\documentclass[12pt,a4paper]{article}
\pdfoutput=1
\usepackage[dvipsnames]{xcolor}
\usepackage{jheppub}
\usepackage{adjustbox}
\usepackage{graphicx}
\usepackage{colortbl}
\usepackage{color}
\usepackage{amsmath}
\usepackage{hyperref}
\usepackage{multirow}
\usepackage{pifont}
\usepackage[normalem]{ulem}
\definecolor{nicered}{rgb}{0.6,0,0}
\definecolor{nicegreen}{rgb}{0.1,0.5,0.1}
\definecolor{niceblue}{rgb}{0,0.4,0.8}
\hypersetup{colorlinks,citecolor= nicegreen,linkcolor=
  nicered,urlcolor=niceblue}

\definecolor{LightCyan}{rgb}{0.88,1,1}
\definecolor{lavenderblush}{rgb}{1.0, 0.94, 0.96}
\definecolor{aliceblue}{rgb}{0.94, 0.97, 1.0}
\definecolor{airforceblue}{rgb}{0.36, 0.54, 0.66}
\definecolor{antiquewhite}{rgb}{0.98, 0.92, 0.84}
\definecolor{outrageousorange}{rgb}{1.0, 0.43, 0.29}
\definecolor{dartmouthgreen}{rgb}{0.05, 0.5, 0.06}
\definecolor{mintcream}{rgb}{0.96, 1.0, 0.98}
\definecolor{vdrgreen}{rgb}{0.0, 0.7, 0.0}

\allowdisplaybreaks

\title{Axionlike particles searches in reactor experiments}
\author[a,b]{D. Aristizabal Sierra,}%
\emailAdd{daristizabal@ulg.ac.be}%
\affiliation[a]{Universidad T\'ecnica Federico Santa Mar\'{i}a -
  Departamento de F\'{i}sica Casilla 110-V, Avda. Espa\~na 1680,
  Valpara\'{i}so, Chile}%
\affiliation[b]{IFPA, Dep. AGO, Universit\'e de Li\`ege, Bat B5, Sart
  Tilman B-4000 Li\`ege 1, Belgium}%
\author[c]{V. De Romeri,}%
\emailAdd{deromeri@ific.uv.es}%
\affiliation[c]{Instituto de F\'{i}sica Corpuscular,
  CSIC/Universitat de Val\`encia, Calle Catedr\'atico Jos\'e
  Beltr\'an, 2 E-46980 Paterna, Spain}%
\author[d]{L. J. Flores,}%
\emailAdd{luisjf89@fisica.unam.mx}%
\affiliation[d]{Instituto de F\'isica, Universidad Nacional Aut\'onoma de M\'exico, A.P. 20-364, Ciudad de M\'exico 01000, M\'exico}%
\author[e]{D.K. Papoulias}%
\emailAdd{d.papoulias@uoi.gr}%
\affiliation[e]{Department of Physics, University of Ioannina
GR-45110 Ioannina, Greece}
\abstract{  
  Reactor neutrino experiments provide a rich environment for the
  study of axionlike particles (ALPs). Using the intense photon flux
  produced in the nuclear reactor core, these experiments have the
  potential to probe ALPs with masses below 10 MeV.  We explore the
  feasibility of these searches by considering ALPs produced through
  Primakoff and Compton-like processes as well as nuclear transitions.
  These particles can subsequently interact with the material of a
  nearby detector via inverse Primakoff and inverse Compton-like
  scatterings, via axio-electric absorption, or they can decay into
  photon or electron-positron pairs.  We demonstrate that
  reactor-based neutrino experiments have a high potential to test
  ALP-photon couplings and masses, currently probed only by
  cosmological and astrophysical observations, thus providing
  complementary laboratory-based searches.  We furthermore show how
  reactor facilities will be able to test previously unexplored
  regions in the $\sim$MeV ALP mass range and ALP-electron couplings
  of the order of $g_{aee} \sim 10^{-8}$ as well as ALP-nucleon
  couplings of the order of $g_{ann}^{(1)} \sim 10^{-9}$, testing
  regions beyond TEXONO and Borexino limits.

}

\date{}
\begin{document}
\maketitle
% -----------
% Section I
% -----------
\section{Introduction}
\label{sec:intro}
In recent years there has been an increasing interest in weakly/feebly
coupled light physics. Arguably the main motivation has been given by
searches at the LHC, which have provided null results of new physics
at the TeV scale. Weakly coupled hidden sectors are a rather plausible
explanation for the absence of such signals. They are motivated as
well by the fact that the absence of dark matter direct detection
signals can be justified if the new secluded sector involves sub-MeV
degrees of freedom. A common feature of these scenarios is the
presence of light scalars/pseudoscalars, usually called axionlike
particles (ALPs), with masses and couplings to Standard Model (SM)
particles which can span over many orders of magnitude.  Examples of
the latter are QCD
axions~\cite{Peccei:1977hh,Peccei:1977ur,Weinberg:1977ma,Wilczek:1977pj},
a widely explored extension of the SM addressing the strong CP
problem. They can also arise as a consequence of spontaneously broken
approximate symmetries, of which examples include familons and
Majorons~\cite{Davidson:1981zd,Wilczek:1982rv,Chikashige:1980ui}.  In these scenarios,
ALPs are the pseudo-Nambu-Goldstone bosons of theories where a
symmetry group $G\supset SU(3)\times SU(2)\times U(1)$ is
spontaneously broken at a high scale, leaving behind light
(pseudo)scalar states.

ALPs may generate signals in a large variety of experimental
environments. They can as well affect the evolution of stars and of
core-collapse supernova explosions and can potentially affect the early
universe dynamics. They can, for instance, contribute to the number of
relativistic degrees of freedom or they can release entropy in the heat
bath thus disrupting BBN predictions (see
e.g.~\cite{Irastorza:2018dyq,DiLuzio:2020wdo} and references
therein). In the high-mass regime, and at the laboratory level, ALPs
have been searched for in beam-dump experiments as well as in
radiative $\Upsilon$ decays at CLEO and BaBar~\cite{Krasny:1987eb,Dobrich:2017gcm,Riordan:1987aw,Bjorken:1988as,Antreasyan:1990cf,
  Aubert:2008as}. They have been as well investigated at accelerator
experiments through different event topologies, first at LEP and more
recently at the LHC~\cite{Abbiendi:2000hh,Aaboud:2017bwk,Bauer:2017ris,Bauer:2018uxu}. Prospects of
exploring regions not yet tested in this mass regime include Belle II,
NA62, NA64, SeaQuest, MATHUSLA, CODEX-b, FASER and SHiP~\cite{Abe:2010gxa,NA62:2017rwk,Gardner:2015wea,Berlin:2018pwi,
  Chou:2016lxi,Curtin:2017izq,Aielli:2019ivi,Anelli:2015pba,Banerjee:2020fue}.

ALPs have been searched for as well in reactor experiments, motivated
by theoretical arguments pointed out first by Weinberg and
subsequently by Donnelly, Freedman, Lytel, Peccei and
Schwartz~\cite{Weinberg:1977ma,Donnelly:1978ty}\footnote{These
  arguments were applicable to the QCD axion and so these
  searches. The experimental results, however, hold as well for
  ALPs.}. To the best of our knowledge, ALP reactor searches looking
for various nuclear magnetic transitions were pioneered  at the
Bugey reactor~\cite{Cavaignac:1982ek}, and followed by searches carried out by
Zehnder and Lehmann et. al. in which instead ALPs were searched for in
nuclear magnetic transitions of $^{127}\text{Ba}^*$ and
$^{65}\text{Cu}^*$~\cite{Zehnder:1981qn,Lehmann:1982bp}. Most
recently, the TEXONO collaboration extended the analysis of
Ref.~\cite{Cavaignac:1982ek} through a comprehensive investigation at
the Kuo-Sheng reactor nuclear power plant, by also employing nuclear
magnetic transitions along with inverse Compton-like and Primakoff
scattering for detection~\cite{Chang:2006ug}.

At present, there is a rapidly developing physics program aiming at
measuring coherent elastic neutrino-nucleus scattering (CE$\nu$NS)
using reactor neutrinos, which offers exciting possibilities to
explore physics beyond the SM. To this purpose, several new
experiments with sub-keV threshold capabilities are in preparation in
The Americas, Western and Eastern Europe and Asia, including---but not
limited to---CONNIE, MINER, CONUS, $\nu$-cleus, Ricochet, RED-100 and
TEXONO~\cite{Aguilar-Arevalo:2019jlr,Agnolet:2016zir,Hakenmuller:2019ecb,Strauss:2017cuu,
  Billard:2016giu,Akimov:2017hee,Wong:2016lmb} (for a detailed review
see Ref.~\cite{Akimov:2019wtg}). Apart from CE$\nu$NS measurements,
this program is aiming at improving upon previous measurements of the
weak mixing-angle, the root-mean-square radii of neutron distributions
for different isotopes, as well as to look for new physics signatures
in the form of e.g. neutrino magnetic moments, light vector and scalar
mediators and neutrino generalized interactions (see
e.g.~\cite{Miranda:2019skf,Giunti:2019xpr,Miranda:2019wdy,Lindner:2016wff,Farzan:2018gtr,
  AristizabalSierra:2018eqm,Liao:2017uzy,AristizabalSierra:2017joc,AristizabalSierra:2019ykk,
  AristizabalSierra:2019ufd,AristizabalSierra:2019zmy,
  Papoulias:2019xaw,Flores:2020lji}). Recently, it has been pointed
out that these technologies also open up a new direction for ALP
searches~\cite{Dent:2019ueq}, further motivating our present work.

In this paper, we investigate the feasibility and extent of such
searches by including leading ALP production and detection mechanisms
at reactor neutrino experiments.  In addition to nuclear magnetic
transitions, considered by previous reactor searches, here we take
into account additional ALP production mechanisms such as Compton-like
and Primakoff scattering. We furthermore consider detection through
their inverse processes as well as via axio-electric absorption and
ALP decays to diphoton and electron pairs.

Our analysis follows minimal ALP assumptions in the sense that it
considers only those processes that allow production and detection
through the same ALP-SM coupling, the exception being production through
nuclear de-excitation. In this case we consider detection through
either ALP-photon or ALP-electron couplings, following to a large
extent the analysis carried out by TEXONO in
Ref.~\cite{Chang:2006ug}. Rather than sticking to a particular
CE$\nu$NS reactor-based experiment, our analysis is formulated as
generic as possible, and therefore being representative of what
current or near-future technologies could achieve.

Due to kinematic constraints, reactor-based searches will cover
regions up to ALP masses of order of $10\,$MeV. Thus they will not cover
some of the regions that Belle II, NA62, NA64, SeaQuest, MATHUSLA, CODEX-b,
FASER and SHiP will have access to. However they can potentially
provide valuable information on regions yet unexplored by laboratory
experiments, while in some cases they can actually explore regions so
far not tested at all. Since CE$\nu$NS reactor experiments will run anyway,
an ALP research program will serve towards widening the physics reach
of these facilities in a direction complementary to other ALP search
strategies.

The remainder of this paper is organized as follows. In
Sec. \ref{sec:models} we present the interactions our analysis deals
with and we briefly discuss some representative QCD axion models. In
Sec. \ref{sec:mechanisms} we discuss the ALPs production and detection
mechanisms considered in the present study, relevant at reactor
experiments. In Sec. \ref{sec:discovery-reach} we present our main
results, the sensitivities achievable in ongoing or near-future
experiments, along with sensitivities expected at the next-generation
CE$\nu$NS reactor experiments. Finally, in Sec. \ref{sec:conclusions}
we summarize and present our conclusions.
% -----------
% Section II
% -----------
\section{Axionlike particle interactions}
\label{sec:models}
Axionlike particles (ALP) are pseudoscalars that feebly
couple to the SM particles. They arise in a variety of new physics
scenarios, including those related with dark matter in which the
pseudoscalar mediates interactions between the dark and visible
sectors~\cite{Nomura:2008ru,Batell:2009di}. In general, they can be
expected in scenarios where the spontaneous breaking of a global symmetry
takes place. In that regard, the Majoron could be thought as a
prototypical example linked to spontaneous lepton number symmetry
breaking, and ultimately with the origin of Majorana neutrino
masses~\cite{Schechter:1981cv}. From a more theoretical point of view,
they can be as well understood as low energy manifestations of string
theories~\cite{Arvanitaki:2009fg,Cicoli:2012sz}.

Strictly speaking, the SM degrees of freedom to which the ALPs couple
depend on the specific model realization. However, from a purely
phenomenological point of view the ALP couplings can be parameterized in
terms of dimension-five effective operators. Here we consider
couplings to photons, electrons and nucleons, for which the
interactions can be written according to
\begin{align}
  \label{eq:interactions-model-independent}
  \mathcal{L}=
  \mathcal{L}_a
  - \frac{1}{4}\,g_{a\gamma\gamma}\, a F_{\mu\nu}\widetilde{F}^{\mu\nu}
  - ig_{aee}a\,\bar e \gamma_5 e
  - ia\bar n\gamma_5\,\left(g_{ann}^{(0)} + \tau_3 g_{ann}^{(1)}\right)n\ ,
\end{align}
where $n$ refers to the proton-neutron isospin doublet and
$g_{ann}^{(0)}$ ($g_{ann}^{(1)}$) to the isosinglet (isotriplet)
axion-nucleon couplings, while the first term in the Lagrangian involves the
ALP kinetic and mass terms as well as the scalar potential.

It is worth noting that the dimension-five operator nature of the couplings to photons is
evident, while for fermions this is not the case. Indeed, couplings to fermions are dictated by derivative
interactions (see e.g.~\cite{Irastorza:2018dyq,DiLuzio:2020wdo})
\begin{equation}
  \label{eq:ALP-f-int}
  \mathcal{L}_f=\frac{C_{af}}{2f_a}\, \bar f\gamma_5\gamma^\mu f\,\partial_\mu a
  \qquad (f=e,p,n)\ ,
\end{equation}
with $f_a$ the ALP decay constant and $C_{af}$ dimensionless strength
couplings. Operators in the Lagrangian~(\ref{eq:ALP-f-int}) are of dimension
five, but they can be cast into those described in the
Lagrangian of Eq.~(\ref{eq:interactions-model-independent}) with the aid of the
fermion equation of motion
$\bar f \partial_\mu\gamma_5 \gamma^\mu f = - i m_f \bar f \gamma_5 f$,
combined with integration by parts. In terms of the decay constant and
fermion masses, the couplings in the Lagrangian~(\ref{eq:ALP-f-int}) are thus
given by
\begin{equation}
  \label{eq:photon-fermion-couplings-decay-constant}
  g_{a\gamma\gamma}=\frac{\alpha}{2\pi}\frac{C_{a\gamma}}{f_a}\ ,\qquad
  g_{aff}=m_f\frac{C_{af}}{f_a}\ .
\end{equation}
So far the discussion applies to ALPs, which are not related with the
solution to the strong CP problem and for which their mass and decay
constant are not related. Reactor experiments are
best suited to test parameter space regions of ALPs rather than of the
QCD axion (see Sec. \ref{sec:discovery-reach}). Nevertheless, in what
follows we briefly provide a few relations derived in QCD axion models
that will enable comparing the extent at which they could be tested
(for comprehensive reviews see e.g.~\cite{Irastorza:2018dyq,DiLuzio:2020wdo}). The interactions in~(\ref{eq:interactions-model-independent}) along with~(\ref{eq:photon-fermion-couplings-decay-constant}) hold as well for
the QCD axion, but in this case the relation
\begin{equation}
  \label{eq:axion-mass}
  m_a\simeq 5.7\left(\frac{10^{12}\,\text{GeV}}{f_a}\right)\,\mu\text{eV}
\end{equation}
applies. This expression combined with
Eq.~(\ref{eq:photon-fermion-couplings-decay-constant}) leads to
relations which can be mapped into the axion mass-coupling plane in
the form of stripes whose widths depend on the UV completion accounting
for the axion effective Lagrangian, namely
\begin{align}
  \label{eq:qcd-axion-photon-coupl}
  g_{a\gamma\gamma}&=2.0\times 10^{-10}\,C_{a\gamma}
  \,\left(\frac{m_a}{\text{eV}}\right)\,
  \text{GeV}^{-1}\ ,
  \\
  \label{eq:qcd-axion-fermion-coupl}
  g_{aff}&=1.8\times 10^{-7}\,C_{af}
  \,\left(\frac{m_f}{\text{GeV}}\right)
  \,\left(\frac{m_a}{\text{eV}}\right)\ .
\end{align}
The dimensionless parameters in Eqs.~(\ref{eq:qcd-axion-photon-coupl})
and~(\ref{eq:qcd-axion-fermion-coupl}) read (see
e.g. Ref. \cite{DiLuzio:2020wdo})
\begin{align}
  \label{eq:dim-less-couplings-gamma}
  C_{a\gamma}&=\frac{\text{E}}{\text{N}} -1.92\ ,
  \\
  \label{eq:dim-less-couplings-proton}
  C_{a\text{p}}&=-0.47 + 0.88 c^0_u - 0.39 c^0_d - C_{a,\text{sea}}\ ,
  \\
  \label{eq:dim-less-couplings-neutron}
  C_{a\text{n}}&=-0.02 + 0.88 c^0_d - 0.39 c^0_u - C_{a,\text{sea}}\ ,
  \\
  \label{eq:dim-less-couplings-sea}
  C_{a,\text{sea}}&=0.038 c^0_s + 0.012 c_c^0 + 0.009 c_b^0 + 0.0035 c_t^0\ ,
  \\
  \label{eq:dim-less-couplings-electron}
  C_{ae}&=c_e^0 + \frac{3\alpha^2}{4\pi^2}
  \left[
   \frac{\text{E}}{\text{N}}\log\left(\frac{f_a}{m_e}\right)
   -
   1.92\log\left(\frac{\text{GeV}}{m_e}\right)
  \right]\ ,
\end{align}
where E and N correspond to the QCD and electromagnetic anomaly
coefficients and the second term in Eq.~(\ref{eq:dim-less-couplings-electron}) arises from one-loop
corrections. Barring the case E/N=0, the axion couples to (gluons)
photons and electrons ($g_{aee}\neq 0$ even for $c_e^0=0$). The
coefficients $c_{f}^0$ are model dependent and in some
instances are zero at the tree level.

The most renowned \textit{invisible axion} UV completions are the KSVZ
and DFSZ models~\cite{Kim:1979if,Shifman:1979if,Zhitnitsky:1980tq,Dine:1981rt}, or
variants of them.  To compare with our sensitivities we select few of them
as follows: for KSVZ-type models we consider those with
$c_{u_i}^0=c_{d_i}^0=c_e^0=0$ and $\text{E/N}\subset [44/3,5/3]$, which cover
a large variety of hadronic axion models~\cite{DiLuzio:2016sbl}. For the case of DSFZ-type models we instead consider the DFSZ-I and DFSZ-II
realizations, for which the relevant parameters are given by 
\begin{align}
  \label{eq:DFSZ-I-II-models}
  \text{DSFZ-I:}&\;\;\frac{\text{E}}{\text{N}}=\frac{8}{3}\ ,\quad
  c_{u_i}^0=-\frac{1}{3}\cos^2\beta\ ,\quad
  c_{d_i}^0=-\frac{1}{3}\sin^2\beta\ ,\quad
  c_{e_i}^0=-\frac{1}{3}\sin^2\beta\ ,
  \nonumber\\
  \text{DSFZ-II:}&\;\;\frac{\text{E}}{\text{N}}=\frac{2}{3}\ ,\quad
  c_{u_i}^0=-\frac{1}{3}\cos^2\beta\ ,\quad
  c_{d_i}^0=-\frac{1}{3}\sin^2\beta\ ,\quad
  c_{e_i}^0=\frac{1}{3}\cos^2\beta\ ,
\end{align}
with $\tan\beta=v_u/v_d\subset [0.25,170]$
($v=\sqrt{v_u^2+v_d^2}=246\,$GeV).
% -----------
% Section II
% -----------
\section{ALPs production and detection mechanisms at reactors}
\label{sec:mechanisms}
Photons are abundantly produced in nuclear power plants. Processes
responsible for $\gamma$ emission are fission, decay of fission
products, capture processes in fuel and other materials, inelastic
scattering in the fuel, and decay of capture products~\cite{roos:1959}. The
prompt photon flux can be described by
\begin{equation}
  \label{eq:flux}
  \frac{\text{d}\Phi_\gamma}{\text{d}E_\gamma}=\frac{5.8\times 10^{17}}
  {\text{MeV}\cdot\text{sec}}\left(\frac{\text{P}}{\text{MW}}\right)
  e^{-1.1\,E_\gamma/\text{MeV}}\ ,
\end{equation}
an approximation originally derived from an analysis of $\gamma$
radiation in the FRJ-1 research reactor~\cite{bechteler1984} and which
has been recently used in dark photon phenomenological
analyses~\cite{Park:2017prx,Ge:2017mcq}.

\begin{figure}[t]
  \centering
  \includegraphics[scale=0.8]{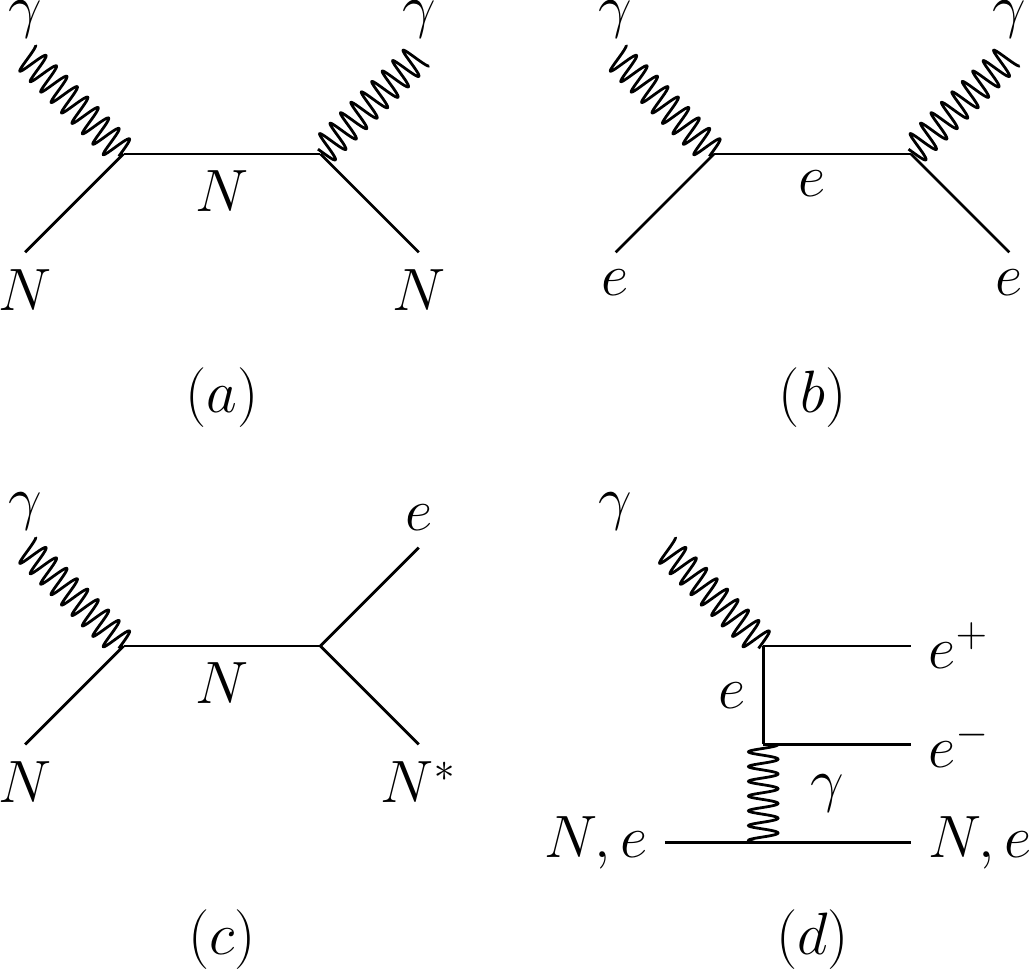}
  \caption{SM photon scattering processes. Process $(a)$ corresponds
    to elastic scattering (Rayleigh), process $(b)$ to inelastic
    scattering (Compton), process $(c)$ to photo-electric absorption
    and process $(d)$ to electron-pair production in the field of the
    nucleus or electron. Photons produced at the nuclear reactor core
    can undergo any of these processes when interacting with the
    reactor fuel.}
  \label{fig:SM-F-diag}
\end{figure}

\begin{figure}[h!]
  \centering
  \includegraphics[scale=0.8]{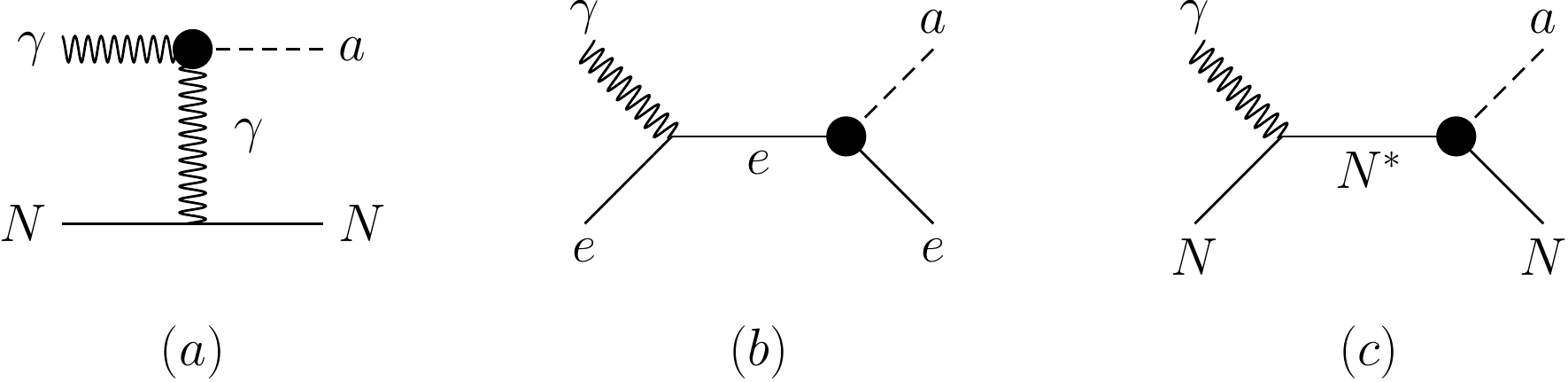}
  \caption{ALPs production mechanisms at a nuclear reactor
    plant. Process $(a)$ corresponds to Primakoff photon-ALP
    conversion, process $(b)$ to Compton-like scattering and process
    $(c)$ to nuclear de-excitation. Depending on the type of SM-ALP
    coupling ($g_{aXX}, X=\gamma,e,n$, where $n$ refers to nucleons), in
    addition to the processes shown in Fig.~\ref{fig:SM-F-diag},
    photons interacting with the reactor fuel can be subject to these
    processes.}
  \label{fig:axion-F-diag}
\end{figure}

Once produced, these photons can interact with the fuel material which
we assume to be $^{235}$U. This choice is motivated by the fact that
most of the reactor experiments our analysis applies to, use this
uranium isotope (totally or partially) as fuel material. Possible
interactions fall in three category processes: (i) a first category
with final-state photons, which involves Rayleigh (elastic) or Compton
(inelastic) scattering; (ii) a second category that accounts for
possible photon absorption by the material, and that involves
photo-electric absorption and $e$-pair production by the nucleus or
electron field, (iii) photon-ALP conversion. Specifically, categories
(i) and (ii) are SM processes, while category (iii) requires a SM-ALP
coupling. The relevant Feynman diagrams for processes (i) and (ii) are
shown in Fig.~\ref{fig:SM-F-diag}. It is worth noting that processes
of type (iii) should involve an initial- and final-state photon and
ALP, respectively. Apart from Primakoff photon-ALP conversion (diagram
$(a)$ in Fig.~\ref{fig:axion-F-diag}), the other two proceed from the
SM Compton and Rayleigh processes by changing the final-state photon
to an ALP (see diagrams $(b)$ and $(c)$ in
Fig.~\ref{fig:axion-F-diag}). The leading mechanisms are Compton-like
production and nuclear de-excitation. All these processes involve
independent SM-ALP couplings: Primakoff production is enabled by
$g_{a\gamma\gamma}$, Compton-like production by $g_{aee}$ and nuclear
de-excitation by $g_{ann}$ ($n=\text{neutron}, \text{proton}$).
\begin{table}
\centering
\renewcommand{\arraystretch}{1.3}
\begin{tabular}{| c | c | c |}
\hline
 \bf{Experiment} & \bf{Nuclear Reactor} & \bf{Power [GW]} \\
\hline   
\hline
\rowcolor{LightCyan}
TEXONO~\cite{Wong:2016lmb}  & Kuo-Sheng Nuclear Power Station    & 2.9 \\ 
\rowcolor{LightCyan}
CONUS~\cite{Hakenmuller:2019ecb}  & Brokdorf   & 3.9 \\ 
\rowcolor{LightCyan}
$\nu$GeN~\cite{Belov:2015ufh}  & Kalinin Nuclear Power Plant     & $\sim 1$ \\   
\rowcolor{LightCyan}
\hline
\rowcolor{lavenderblush}
MINER~\cite{Agnolet:2016zir}  & TRIGA  1    & $10^{-3}$ \\  
\rowcolor{lavenderblush}
$\nu$CLEUS~\cite{Strauss:2017cuu}  & FRM2   & 4 \\
\rowcolor{lavenderblush}
Ricochet~\cite{Billard:2016giu}  & Chooz Nuclear Power Plant    & 8.54 \\
\hline
\rowcolor{aliceblue}
RED-100~\cite{Akimov:2017hee}  & Kalinin Nuclear Power Plant    &  $\sim 1$ \\  
\rowcolor{aliceblue}
SBC~\cite{eric_2020}  & ININ (or Laguna Verde)   &  $10^{-3}$ (1.5) \\  
\hline
\rowcolor{antiquewhite}
CONNIE ~\cite{Moroni:2014wia}  & Angra 2  & 3.8    \\
\rowcolor{antiquewhite}
vIOLETA ~\cite{Fernandez-Moroni:2020yyl} & Atucha II & 2 \\
\hline
\rowcolor{mintcream}
SoLid ~\cite{Abreu:2020bzt}  & BR2    & $(0.4,1)\times 10^{-1}$ \\
 \rowcolor{mintcream}
NEON~\cite{neon}  & Hanbit Nuclear Power Plant   &  2.8   \\
\hline    
\end{tabular}
\caption{Reactor neutrino experiments to which our analysis may apply. 
  We also present the associated nuclear reactor along with its thermal power. The color 
  codes are associated with the detection techniques the experiments rely on as 
  specified in Tab.~\ref{tab:reac_exps_det}.}
\label{tab:reac_exps_prod}
\end{table}

\begin{figure}[h!]
  \centering
  \includegraphics[scale=0.8]{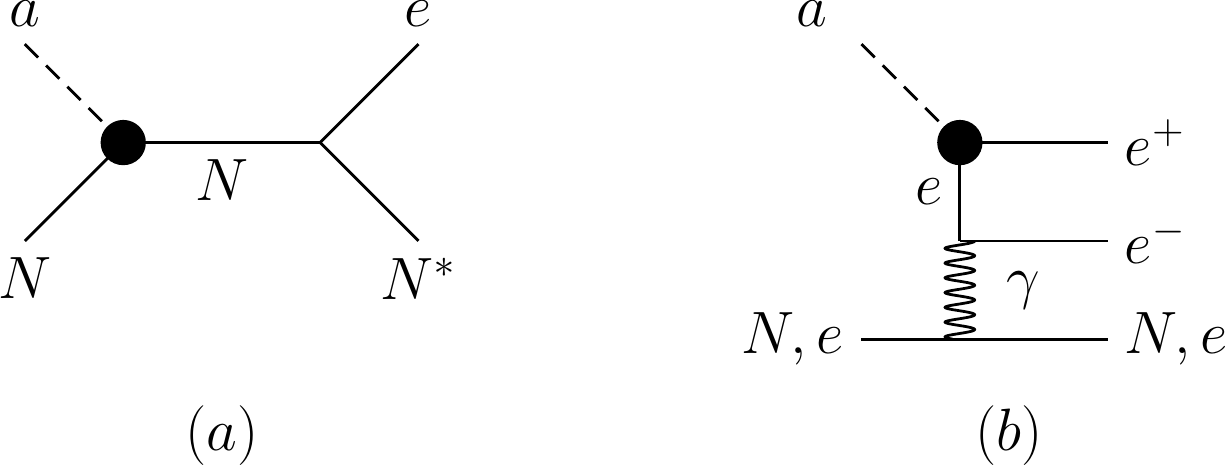}
  \caption{ALPs detection mechanisms, along with inverse Primakoff,
    inverse Compton-like and nuclear excitation (inverse processes in
    Fig.~\ref{fig:axion-F-diag}). Process $(a)$ corresponds to the
    axio-electric process (photo-electric absorption equivalent,
    diagram $(c)$ in Fig.~\ref{fig:SM-F-diag}), while process $(b)$ to
    axion $e-$pair production ($e-$pair production in the field of the
    nucleus or electron equivalent, diagram $(d)$ in
    Fig.~\ref{fig:SM-F-diag}). As in the production case, whether such
    detection mechanisms are operative depends on the type of coupling
    the analysis is subject to. See
    Tab.~\ref{tab:processes-and-couplings} for details.}
  \label{fig:axion-F-diag-det}
\end{figure}
Primakoff, Compton-like and nuclear de-excitation can also serve as
detection mechanisms. The other possible detection processes follow
from the SM photo-electric absorption and $e$-pair production by
changing $\gamma\to a$, as shown in
Fig.~\ref{fig:axion-F-diag-det}. In the former case, axio-electric
absorption~\cite{Avignone:1986vm}, an ALP is absorbed by an atom
resulting in electron emission. In the latter, an ALP is absorbed and
an electron-pair is emitted. Depending on the ALP lifetime, detection
can proceed as well through ALP decay processes: $a\to \gamma\gamma$,
$a\to e^+e^-$ and $a\to nn$  (since for the last two processes
$m_a>2m_e$ and $m_a>2m_n$  should be satisfied, decay to nucleons is
not a viable possibility in nuclear reactor experiments). Table
\ref{tab:processes-and-couplings} shows the different production and
detection mechanisms together with the relevant couplings.
From this table one can see that \textit{self-contained
  one-parameter analyses} (axion mass aside) of ALP production should
be done as follows:
\begin{itemize}
\item Analyses with $g_{a\gamma\gamma}$ should involve production
  through Primakoff and detection through inverse Primakoff and
  decay into a $\gamma-$pair final states.
\item Analyses with $g_{aee}$ should include production through
  Compton-like and detection through inverse Compton-like,
  axio-electric, $e-$pair production in $e$ field and decay into an electron-positron pair final
  states.
\item Analyses with $g_{ann}$ should be done considering production
  through nuclear de-excitation and detection through nuclear
  excitation.
\end{itemize}
We relax our analysis from considering  ALP $e-$pair production in the
field of $N$ or $e^-$ (diagram $(b)$ in
Fig.~\ref{fig:axion-F-diag-det}). Although we do not calculate the ALP
$e-$pair production cross sections, we expect that the detection through these
processes should produce a rather small event yield, similar to their SM counterparts
which are suppressed (see Fig.~\ref{fig:-photon-x-sec}). We do not consider
either ALP nuclear excitation in detection. Hence, in our analysis of
ALP production through nuclear de-excitation
(Sec. \ref{sec:neutron-ALP-coupling})  we consider detection 
mechanisms which involve other ALP couplings (either $g_{aee}$ or $g_{a\gamma\gamma}$).

% ---------------
% Table
% ---------------
\renewcommand{\arraystretch}{1.5}
\begin{table}[t]
  \centering
  \begin{tabular}{|c|c|c|c|c|}\hline
    \rowcolor[gray]{0.8}\multicolumn{5}{|c|}{\textbf{Scattering processes}}\\\hline
    % 1st row
    \multicolumn{2}{|c|}{Process} & Coupling & Prod & Det
    \\\hline
    % 2nd row
    Primakoff & $\gamma+N\leftrightarrow a+N$& \cellcolor{Apricot}$g_{a\gamma\gamma}$&
    \ding{52}&
    \ding{52}
    \\\hline
    % 3rd row
    Compton-like & $\gamma+e^-\leftrightarrow a+e^-$
    & \cellcolor{Salmon}$g_{aee}$& 
    \ding{52}& 
    \ding{52}
    \\\hline
    % 4th row
    Nuclear de-excitation 
    & $\gamma+N\leftrightarrow N^*\to a+N$& 
    \cellcolor{Emerald}$g_{ann}$&
    \ding{52}&
    \ding{52}
    \\\hline
    % 5th row
    Axio-electric & $a+e^-+Z\to e^-+Z$& \cellcolor{Salmon}$g_{aee}$&
    \ding{55}&
    \ding{52}
    \\\hline
    % 6th row
    $e-$pair production in $N$ & $a+N\to e^-+e^-+ N$& 
    \cellcolor{Emerald}$g_{ann}$&
    \ding{55}&
    \ding{52}
    \\\hline
    % 7th row
    $e-$pair production in $e$ 
    & $a+e^-\to e^-+e^+ + e^-$
    & \cellcolor{Salmon}$g_{aee}$
    & \ding{55}
    & \ding{52}
    \\\hline\hline
    % 8th row
    \rowcolor[gray]{0.8}\multicolumn{5}{|c|}{\textbf{Decay processes}}\\\hline
    \multicolumn{2}{|c|}{Process} & Coupling & Prod & Det
    \\\hline
    % 9th row
    $\gamma$-pair final state& $a\to\gamma+\gamma $
    & \cellcolor{Apricot}$g_{a\gamma\gamma}$
    &\ding{55}
    &\ding{52}
    \\\hline
    % 10th row
    $e$-pair final state& $a\to e^-+e^+ $& \cellcolor{Salmon}$g_{aee}$&
    \ding{55}&
    \ding{52}
    \\\hline
    % 10th row
    $n$-pair final state& $a\to n+n $& \cellcolor{Emerald}$g_{ann}$&
    \ding{55}&
    \ding{55}
    \\\hline
\end{tabular}
\caption{ALPs production and detection mechanisms. Check marks (crosses) 
  are used to identify (exclude) production and/or detection processes. 
  Color refers to processes that should be---in principle---included in a 
  self-contained one-parameter analysis. Note that decay to nucleon 
  pairs is forbidden by kinematic arguments. } 
\label{tab:processes-and-couplings}
\end{table}
Phenomenological ALP studies follow a three-step analysis: (i)
production, (ii) propagation towards the detection chamber and (iii)
detection. The relevant production mechanisms assumed in the present work, 
namely, Primakoff conversion, Compton-like scattering
or nuclear de-excitation, generate an ALP flux which propagates and
can potentially reach the detector depending on the ALP lifetime,
determined by its survival probability. Neglecting possible
interactions with the detector shielding material, ALPs reaching the
detector will then interact with the target material yielding---in principle---a measurable signal. The quantities that intervene in
this three-step process then combine through a convolution, that
defines the differential event yield from which the number of counts
at the detector is calculated. 

Possible signals depend on the kind of process ALPs are subject
to. For instance, in the axio-electric process ALP absorption by the
material produces phonons (from the temperature rise of the absorber)
and electron-hole pairs leading to heat and ionization
signals. Ideally, detection through the axio-electric process should
then be done using dielectric crystals, whose properties are optimized
for such type of signatures. 
DAMA, CDMS and more recently the
SuperCDMS Soudan experiment have followed this strategy, i.e. DAMA using NaI crystals,
while CDMS and SuperCDMS Soudan using a cryogenic detector based on
high-purity germanium and silicon crystals
\cite{Ahmed:2009ht,Bernabei:2005ca,Aralis:2019nfa}.
The different
detector technologies as well as detector specifications of typical experiments to which our analysis
can apply are shown in Tab.~\ref{tab:reac_exps_det}. These parameters
motivate the generic numbers used in our analysis: germanium as
material target, $m_\text{det}=10\,$kg and $L=10\,$m (see
Sec. \ref{sec:discovery-reach}).
\begin{table}%[htb!]
\centering
%\begin{adjustbox}{angle=90}
\begin{tabular}{| c | c | c | c | c |}
  \hline
  \bf{Detector} & \bf{Experiment}  & \bf{Material} & \bf{$\text{m}_\text{det}$[kg]}& $\boldsymbol{L}$ \bf{[m]}  \\
\hline   
\hline 
\rowcolor{LightCyan}
 \textcolor{blue}{Semiconductor} &TEXONO~\cite{Wong:2016lmb}  & Ge &  1.06&   28  \\ 
\rowcolor{LightCyan}
\textcolor{blue}{detectors} & CONUS ~\cite{Hakenmuller:2019ecb}  & Ge &  1 &   17.1   \\ 
\rowcolor{LightCyan}
\textcolor{blue}{(ionization)} & $\nu$GeN~\cite{Belov:2015ufh}  & Ge &  1.6-5 &   10-12  \\
\hline
\rowcolor{lavenderblush}
\textcolor{red}{Low}  & MINER~\cite{Agnolet:2016zir}  & Ge, Si  &  4&   1-2.5 \\  
\rowcolor{lavenderblush}
\textcolor{red}{temperature}  & $\nu$CLEUS~\cite{Strauss:2017cuu}  & $\rm CaWO_4$, $\rm Al_2O_3$  &  $10^{-2}$ &   15-100  \\ 
\rowcolor{lavenderblush}
\textcolor{red}{bolometers} & Ricochet~\cite{Billard:2016giu}  & Ge, Zn &  10 &   355/469  \\    
\hline
\rowcolor{aliceblue}
\textcolor{airforceblue}{Liquid noble-gas} & RED-100~\cite{Akimov:2017hee}  & Xe  & 100  &  19  \\  
 \rowcolor{aliceblue}
 \textcolor{airforceblue}{detectors (TPC)} & SBC~\cite{eric_2020}  & LAr, Xe  & 10&  3/30  \\  
\hline
\rowcolor{antiquewhite}                                                                
\textcolor{outrageousorange}{CCD} & CONNIE ~\cite{Moroni:2014wia}  & Si &  $\sim 0.05$ &   30  \\ 
\rowcolor{antiquewhite} 
& vIOLETA ~\cite{Fernandez-Moroni:2020yyl} & Si & 1 & 12 \\
\hline
\rowcolor{mintcream}
\textcolor{dartmouthgreen} {Scintillators} & SoLid ~\cite{Abreu:2020bzt}  & $\rm ^{6}LiF:ZnS(Ag)$  &  1600 &  $\sim 7.6$\\   
 \rowcolor{mintcream}
& NEON~\cite{neon}  & NaI[Tl]  & 3.3-10&  24\\
\hline    
\end{tabular}
%\end{adjustbox}
\caption{Detectors to which our analysis may apply, along with relevant parameters
  for ALP detection. With germanium as a target material, 
  $m_\text{det}=10\,$kg and $L=10\,$m are rather representative of the experiments 
  our analysis can cover. CCD refers to charge-coupled devices.}
\label{tab:reac_exps_det}
\end{table}
\subsection{Production and detection through photon-ALP coupling}
\label{sec:Primakoff}
Let us now focus on Primakoff-like production and
detection processes. In full generality, the determination of the ALP flux
involves a convolution (over ingoing photon energies) of the photon
flux and the photon-ALP conversion differential cross section,
properly normalized to account for the processes (i) and (ii). With the
ALP flux at hand, and after assuring the ALP survival probability is
sizeable enough, the detection proceeds through a second convolution (over
ingoing ALP energies) of the ALP flux and the ALP-photon conversion
differential cross section. If the ALP undergoes decay within the
detector chamber, photon signals can be triggered as well by its
decay. In such a scenario the ALP detection is therefore a combination of
inverse-Primakoff scattering and decay contributions, with the latter being
mainly relevant in the ``high'' ALP mass region (the decay partial
width is proportional to the ALP mass). Let us discuss this in more
detail.

\begin{figure}
  \centering
  \includegraphics[width = 0.7 \textwidth]{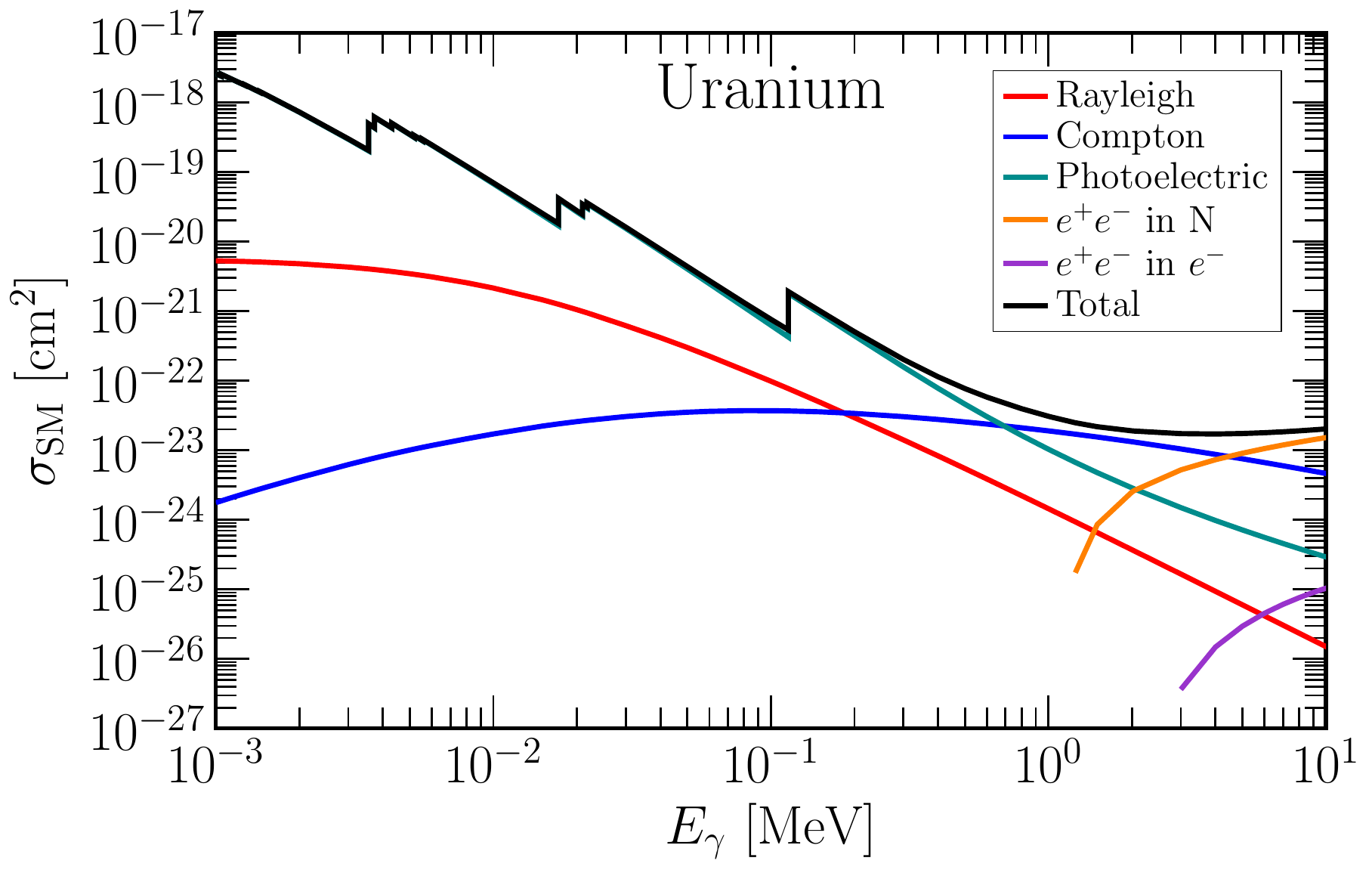}
  \caption{SM photon cross sections for uranium taken from the XCOM
    photon cross section database~\cite{xcom:2010}.}
  \label{fig:-photon-x-sec}
\end{figure}

Accounting for the probability that an ALP of energy $E_a$ is emitted
after a $\gamma-^{235}\text{U}$ interaction, the ALP flux proceeds
then from the following convolution
\begin{equation}
  \label{eq:ALPflux-Primakoff}
  \frac{d\Phi^\text{P}_a}{dE_a}= \mathcal{P}_\text{surv}
  \int_{E_{\gamma^\prime,\text{min}}}^{E_{\gamma^\prime,\text{max}}}
  \frac{1}{\sigma_\text{Tot}}
  \frac{d\sigma^\text{P}_\text{P}}{dE_a}(E_{\gamma^\prime},E_a)
  \,\frac{d\Phi_{\gamma^\prime}}{dE_{\gamma^\prime}}  
  \,\, dE_{\gamma^\prime}\ ,
\end{equation}
where $\sigma_\text{Tot}=\sigma_\text{SM}+\sigma^\text{P}_\text{P}$,
with the SM cross section being
$\sigma_\text{SM}=\sigma_\text{R}+\sigma_\text{C}+\sigma_\text{PE}+\sigma_{e-\text{pair}}$,
while $\sigma^\text{P}_P\equiv\sigma_{\gamma\to a}$ is the total
Primakoff scattering cross section for ALP production.  Results for
the SM components are obtained from the XCOM: Photon cross section
database~\cite{xcom:2010} and shown in Fig.~\ref{fig:-photon-x-sec}
for completeness. The integration limits are determined by the
kinematic relation~\footnote{To avoid confusion, throughout the
  manuscript the incident and outgoing photon energies are denoted by
  $E_{\gamma^\prime}$ and $E_{\gamma}$, respectively.}
\begin{equation}
  \label{eq:Egamma-Ea-relation}
  E_{\gamma^\prime}=\frac{2E_a M_N - m_a^2}{2(M_N - E_a + |\vec p_a|\cos\theta)}\ ,
\end{equation}
with $E_{\gamma^\prime}^\text{min}$ ($E_{\gamma^\prime}^\text{max}$)
found for $\cos\theta=+1$ ($\cos\theta=-1$) and $M_N$ being the
nuclear mass. Differences between the upper and lower limits are in the
sub-keV ballpark, almost independent of the ALP
mass. Kinematically, this implies that a photon with a given
energy $E_{\gamma^\prime}$ produces ALPs within a rather narrow energy
range.

The ALP survival probability, assuring that the ALP flux reaches the
detector, can be cast as~\cite{Dent:2019ueq}
\begin{equation}
  \label{eq:survival}
  \mathcal{P}_\text{surv}=e^{-L E_a/|\vec{p}_a|\tau}\ ,
\end{equation}
where $L$ refers to the distance between the reactor and detector and
$\tau$ to the ALP lifetime in the fixed target frame, determined in
turn by the ALP total decay width:
$\tau^{-1}=\Gamma_{a\to 2\gamma}\times m_a/E_a$, while the decay width
reads
\begin{align}
  \label{eq:decay-widths}
  \Gamma_{a\to 2\gamma}\equiv \Gamma(a\to \gamma\gamma)=
  \frac{g_{a\gamma\gamma}^2m_a^3}{64\pi}\ .
\end{align}

Thus, for a photon with energy $E_{\gamma^\prime}$ the consequent ALP
distribution can be approximated with a monochromatic
one~\footnote{For definiteness, in what follows the cross sections are
  written in the form $\sigma_{X}^{Y}$, where $X=$(P, D) indicates the
  corresponding mechanism e.g. production or detection and $Y=$(P, C,
  A) indicates the nature of the process e.g. Primakoff-like,
  Compton-like and axio-electric, respectively. Similarly, the
  ALP flux is given in the form
  $\frac{\text{d}\Phi_a^Z}{\text{d}E_a}$, where $Z=$(P, C, MT)
  accounts for Primakoff-, Compton- and Magnetic Transition
  (MT)-induced ALPs, while the number of events  is written in the
  form $ \mathcal{N}^Z_Y$.}
\begin{equation}
  \label{eq:approximation-xsec}
  \frac{\text{d}\sigma^\text{P}_P}{\text{d}E_a}\simeq 
  \sigma^\text{P}_\text{P}(E_a)\delta(E_{\gamma^\prime}-E_a)\ .
\end{equation}
This observation allows to determine the ALP flux with a
simplified version of Eq.~(\ref{eq:ALPflux-Primakoff}),
namely 
\begin{equation}
  \label{eq:axion-flux-simplified}
  \frac{\text{d}\Phi^\text{P}_a}{\text{d}E_a}=\frac{\sigma^\text{P}_\text{P}}{\sigma_\text{Tot}}
  \left .\frac{d\Phi_{\gamma^\prime}}{dE_{\gamma^\prime}}\right|_{E_{\gamma^\prime}=E_a}\ .
\end{equation}

For the calculation of the total Primakoff scattering cross section
we rely on the expression~\cite{Aloni:2019ruo}
\begin{align}
  \label{eq:Primakoff-x-sec}
  \frac{\text{d}\sigma^\text{P}_\text{P}}{\text{d}t}=
  2\alpha Z^2F^2(t)\,g_{a\gamma\gamma}^2\frac{M_N^4}{t^2(M_N^2-s)^2(t-4 M_N^2)^2}
  &\left\{m_a^2t(M_N^2 + s) - m_a^4 M_N^2 \right .
  \nonumber\\
  &\left . - t\left[(M_N^2-s)^2 + st\right]\right\}\ ,
\end{align}
a result that does not
involve any approximation (see discussion below).
Here $s,t$ are Mandelstam variables and $\alpha Z^2F^2(t)$ is the effective electric charge to which the
photon couples to. Its value is controlled by $F^2(t)$, a form factor
that we normalize to unity. For a sufficiently small exchanged momentum
$|q|=\sqrt{-t}\leq \sqrt{7.39 m_e^2}\simeq 7.0\times
10^{-3}\,\text{fm}^{-1}$,
the DeBroglie wavelength of the scattering process is large enough so
to partially involve the atomic electron cloud~\cite{Tsai:1986tx}. In
that case the effective electric charge is diminished (screened), with
the reduction depending on how small the exchanged momentum
becomes. For such regime we use the following atomic form factor
\cite{Tsai:1973py}
\begin{equation}
  \label{eq:atomic-FF}
  F_\text{Atomic}^2(t)=\left(\frac{a_0^2t}{1 - a_0^2 t}\right)^2\ ,
\end{equation}
with $a_0$ calculated in the Thomas-Fermi atom using the Moliere
representation
\begin{equation}
  \label{eq:a-parameter-Atomic-FF}
  a_0 = \frac{184.15}{\sqrt{2.718}}\frac{1}{Z^{1/3}m_e}\ .
\end{equation}
Since we choose $t=q^2<0$, contrary to what Ref.~\cite{Tsai:1973py}
does, Eq.~(\ref{eq:atomic-FF}) results from shifting $t\to -t$ in the
form factor provided by that reference. For a larger exchanged momentum
$|q|=\sqrt{-t}> \sqrt{7.39 m_e^2}\simeq 7.0\times
10^{-3}\,\text{fm}^{-1}$
the DeBroglie wavelength becomes smaller and hence the atomic electron cloud
is not relevant anymore. However as $q$ increases the nuclear electric
charge diminishes as well, therefore the atomic form factor has to be
replaced by a nuclear form factor which we choose to be of the Helm
form~\cite{Helm:1956zz}\footnote{Ref.~\cite{Aloni:2019ruo} uses the
  two-parameter Fermi model form factor. Differences between results
  generated using this choice or the one we have adopted are
  negligible~\cite{AristizabalSierra:2019zmy}.}. It is worth
emphasizing that while the inclusion of a nuclear form factor
has a negligible effect on our results, the inclusion of the atomic
form factor is instead important, in particular in the low ALP mass region
$m_a\lesssim 10^4\,$eV. This can be understood from the nature of the
Primakoff process, for which the largest amount of events are found at
small $t$ or along the forward direction.

The calculation of the Primakoff cross section in
Ref.~\cite{Tsai:1986tx} is valid in the forward scattering limit as
well as in the limit $E_\gamma=E_a$ and $m_a\to 0$. In contrast, the
result in Eq.~(\ref{eq:Primakoff-x-sec}) does not involve any of those
approximations, even if $E_\gamma=E_a$ is still valid to a large degree, as we have
already pointed out.  However, departures from forward scattering as
well as a massive ALP can produce sizable deviations, particularly
relevant at photon energy threshold\footnote{We thank Bhaskar Dutta
  and Adrian Thompson for discussions on this subject.}.

For the integration of the Primakoff differential cross section in Eq.~(\ref{eq:approximation-xsec}) we use the following
limits~\cite{Tanabashi:2018oca}
\begin{equation}
  \label{eq:tmin-tmax}
  t_\text{max}(t_\text{min})=\frac{m_a^4}{4s} - (p_{1\text{cm}}\mp
  k_{1\text{cm}})^2\ ,
\end{equation}
where $p_1$ ($k_1$) refer to the initial-state photon (final-state
axion) three-momentum given by
\begin{equation}
  \label{eq:3-mom-P}
  p_{1\text{cm}}=\frac{s-M_N^2}{2\sqrt{s}}\ ,
  \qquad
  k_{1\text{cm}}=
  \left[
    \frac{(s+m_a^2-M_N^2)^2}{4s} - m_a^2
  \right]^{1/2}\ ,
\end{equation}
with $s=M_N^2 + 2 M_N E_{\gamma^\prime}$.

Then, the differential event rate consists of two terms, one from the inverse-Primakoff
scattering and another one from ALP decays $a\to \gamma\gamma$
\begin{equation}
  \label{eq:diff-event-rate}
  \frac{\text{d}\cal{N}^\text{P}_\text{Tot}}{\text{d}E_a}=
  \frac{\text{d}\cal{N}^\text{P}_\text{P}}{\text{d}E_a}
  +
  \frac{\text{d}\cal{N}^\text{P}_\text{decay}}{\text{d}E_a}\ .
\end{equation}
Here, the first term results from a convolution of the ALP flux and the
ALP-photon differential cross section\footnote{The inverse Primakoff
  differential cross section (differential cross section in detection, $\text{d}\sigma^\text{P}_D/\text{d}E_\gamma$)
  is twice as large as the one in production given in Eq.~(\ref{eq:Primakoff-x-sec}).}, namely
\begin{equation}
  \label{eq:diff-event-rate-scattering-Primakoff}
  \frac{\text{d}\cal{N}^\text{P}_\text{P}}{\text{d}E_a}=m_\text{det}\frac{N_T \Delta t}{4\pi L^2}\int
  \frac{\text{d}\Phi^\text{P}_a}{\text{d}E_a}\,\frac{\text{d}\sigma^\text{P}_D}{\text{d}E_\gamma},\text{d}E_\gamma\ ,
\end{equation}
which taking into account Eq.~(\ref{eq:approximation-xsec}) allows
writing the ALP yield through Primakoff scattering according to
\begin{equation}
  \label{eq:ALP-yield-P}
  {\cal N}^\text{P}_\text{P}=m_\text{det}\frac{N_T \Delta t}{4\pi L^2}
  \int_{E_a^\text{min}}^{E_a^\text{min}}
  \sigma^\text{P}_P(E_a)
  \,\frac{\text{d}\Phi^\text{P}_a}{\text{d}E_a}\,
  \,\text{d}E_a\ ,
\end{equation}
where $N_T=N_A/m_\text{molar}$ with
$N_A=6.022\times 10^{23}/\text{mol}$ and $m_\text{molar}$ the target
nuclei molar mass in $\text{kg}/\text{mol}$ ($N_T$ thus measures the
amount of nuclei/kg). Moreover, $\Delta t$ stands for the data-taking time and $m_\text{det}$ refers to the
detector mass in kg.  Integration limits are $E_a^\text{min} =25$~keV
and $E_a^\text{max}=10\,$MeV (determined by the
photon natural threshold at a reactor). The second term in Eq.~(\ref{eq:diff-event-rate})
is instead given by
\begin{equation}
  \label{eq:decay}
  \frac{\text{d}{\cal N}^\text{P}_\text{decay}}{\text{d}E_a}=\frac{\mathcal{A} \Delta t}{4\pi L^2}\frac{\text{d}\Phi^\text{P}_a}{\text{d}E_a}
  \mathcal{P}_\text{decay}\ ,
\end{equation}
where $P_\text{decay}$ accounts for the probability that the decay occurs within the
detector, and reads
\begin{equation}
\mathcal{P}_\text{decay}=
  1 - e^{-L_\text{det} E_a/|\vec{p}_a|\tau}\ ,
\end{equation}
where $L_\text{det}$ stands for the detector length and $\mathcal{A} = L_\text{det}^2 $ denotes the detector transverse area.

%%%%%%%%%%%%%%%%%%%%%%%%%%%%%%%%%%%%%%%%%%%%%%%%
\subsection{Production and detection through electron-ALP coupling}
\label{sec:electron-ALP-coupling}

The MeV photons produced in a reactor may also scatter off electrons in the materials of the reactor core and they can produce
ALPs via the Compton-like process $\gamma e^- \rightarrow a e^-$. This process requires the individual SM-ALP coupling $g_{aee}$ to be non-zero. The ALPs so produced can be subsequently detected via three possible processes relying on the same $g_{aee}$ coupling: inverse Compton-like, axio-electric and decay into $e^+ e^-$ pairs. The total number of observed ALP events is computed similarly to the Primakoff case.

As a first step, we calculate the Compton-produced ALP flux via a convolution of the reactor photon flux $d\Phi_{\gamma^\prime}/dE_{\gamma^\prime}$ and the production cross section $d\sigma_\text{P}^\text{C}/dE_a$, as

\begin{equation}
  \label{eq:ALPflux-Compton}
  \frac{\text{d}\Phi^\text{C}_a}{\text{d}E_a}= \mathcal{P}_\text{surv} \int_{E_{\gamma^\prime,\text{min}}}^{E_{\gamma^\prime,\text{max}}}
  \frac{1}{\sigma_\text{Tot}}
  \frac{\text{d}\sigma^\text{C}_\text{P}}{\text{d}E_a}(E_{\gamma^\prime},E_a)\,\frac{\text{d}\Phi_{\gamma^\prime}}{\text{d}E_{\gamma^\prime}} \, \, \text{d}E_{\gamma^\prime}\ ,
\end{equation}
where we assume $E_{\gamma^\prime,\text{min}} = (2 m_e m_a + m_a^2)/(2 m_e)$, $E_{\gamma^\prime,\text{max}} = 10$ MeV and  $\sigma_\text{Tot}=\sigma_\text{SM}+\sigma^\text{C}_\text{P}$.  The differential Compton-like cross section for ALP production has threshold $s> (m_a + m_e)^2$ and is given by~\cite{Brodsky:1986mi}

\begin{equation}
\frac{\text{d}\sigma^\text{C}_\text{P}}{\text{d}E_a} = \frac{Z \pi g_{aee}^2 \alpha x}{4 \pi (s - m_e^2)(1 - x) E_{\gamma^\prime}} \left[x - \frac{2 m_a^2 s}{(s - m_e^2)^2} + \frac{2 m_a^2}{(s -m_e^2)^2} \left(\frac{m_e^2}{1- x} + \frac{m_a^2}{x}\right)\right] \, ,
\end{equation}
where $s$ is the Mandelstam variable and 
\begin{equation}
x= 1 - \frac{E_a}{E_{\gamma^\prime}} + \frac{m_a^2}{2 E_{\gamma^\prime} m_e} \, .
\end{equation}
The total Compton-like production cross section is obtained by integrating $\text{d}\sigma_\text{P}/\text{d}E_a$ over the integration limits~\cite{Chanda:1987ax}: 
$$x_\text{min (max)} = \frac{1}{2 s (s - m_e^2)} [(s - m_e^2) (s - m_e^2 + m_a^2) \mp (s - m_e^2) \sqrt{(s -m_e^2 + m_a^2)^2 - 4 s m_a^2}].$$
One then gets the differential number of ALP events with a second convolution, of the ALP flux $\text{d}\Phi_a/\text{d}E_a$ with the detection cross section. The latter will be different depending on the detection mechanism: 

\begin{align}
  \label{eq:diff-event-rate-Compton}
  &\frac{\text{d}\cal{N}^\text{C}_\text{C}}{\text{d}E_\gamma}=m_\text{det}\frac{N_T \Delta t}{4\pi L^2}\int_{E_{a,\text{min}}}^{E_{a,\text{max}}}
    \frac{\text{d}\Phi^\text{C}_a}{\text{d}E_a}\,\frac{\text{d}\sigma^\text{C}_\text{D}}{\text{d}E_\gamma}(E_{\gamma},E_a)\,\,\text{d}E_a, \\
  &\frac{\text{d}\cal{N}^\text{C}_\text{A}}{\text{d}E_a}=m_\text{det}\frac{N_T \Delta t}{4\pi L^2}
  \frac{\text{d}\Phi^\text{C}_a}{\text{d}E_a}\,\sigma^\text{A}_\text{D}(E_{\gamma},E_a), \\
 &\frac{\text{d}\cal{N}^\text{C}_\text{decay}}{\text{d}E_a}=\frac{\mathcal{A}  \Delta t}{4\pi L^2}
  \frac{\text{d}\Phi^\text{C}_a}{\text{d}E_a}\,\mathcal{P}_\text{decay},
\end{align}
namely inverse Compton-like, axio-electric, and $e^+ e^-$ decay, respectively. In this case, the total ALP decay width reads: $\tau^{-1}=\Gamma_{a\to e^+ e^-}\times m_a/E_a$, where the $a \to e^+ e^-$ decay
width is
\begin{equation}
  \label{eq:decay-width-ee}
  \Gamma_{a\to e^+ e^-} = \frac{g_{aee}^2m_a}{8\pi}\sqrt{1-4\frac{m_e^2}{m_a^2}}\, ,
  \end{equation}
while, the relevant detection cross sections can be cast as~\cite{Avignone:1988bv,Dimopoulos:1986mi,Pospelov:2008jk,Derevianko:2010kz}
\begin{align}
  \label{eq:det-xsecs-gaee}
 \frac{\text{d}\sigma^\text{C}_\text{D}}{\text{d}E_\gamma} &= \frac{Z g_{aee}^2 \alpha E_\gamma}{4 m_e^2 |\vec p_a|}  \left| \frac{2 (E_a + m_e- |\vec p_a| \text{cos}\theta)^2}{|\vec p_a| y} \right| \, \nonumber\\
 &\times\left(1 + \frac{4 m_e^2 E_\gamma^2}{y^2} - \frac{4 m_e E_\gamma}{y} - \frac{4 m_a^2 |\vec p_a|^2 m_e E_\gamma (1 - \text{cos}^2\theta)}{y^3}\right) \, \nonumber\\
 \sigma^\text{A}_\text{D} &= \frac{g_{aee}^2}{\beta} \frac{3 E_a^2}{16 \pi \alpha m_e^2} \left(1 - \frac{\beta^{2/3}}{3}\right) \sigma_{\text{PE}}\, ,
\end{align}
where $\sigma_\text{PE}$ denotes the SM photoelectric cross section of the detector material.
Here, 
$\theta$ is the scattering angle, $y = 2 m_e E_a + m_a^2$ and $\beta = |\vec p_a|/E_a$~\footnote{This definition of $\beta$ should not be confused with the one given in Eq.~(\ref{eq:DFSZ-I-II-models}).}. The integration limits $E_{a,\text{min (max)}}$ are functions of $E_\gamma$: 

\begin{align}
\label{eq:Eaminmax}
E_{a,\text{min}} &= \frac{(2 E_\gamma m_e - m_a^2)(E_\gamma - m_e) + E_\gamma\sqrt{ (m_a^2 + 2 E_\gamma m_e )^2 - 4 m_a^2 m_e^2}}{2 (2 E_\gamma m_e - m_e^2)}\, \\
E_{a,\text{max}} &= E_{\gamma^\prime,\text{max}} (1 - x_\text{min}) + m_a^2/(2 m_e) \,.
\end{align}
Finally, the total number of observed ALP events from Compton-like
production in a reactor neutrino experiment would be
\begin{align}
  \label{eq:total-events-Compton}
  \cal{N}^\text{C}_\text{C}&=\int_{E_{\gamma, \text{min}}}^{E_{\gamma, \text{max}}} \frac{\text{d}\cal{N}^\text{C}_\text{C}}{\text{d}E_\gamma} \, \text{d}E_\gamma\nonumber \\
  &= m_\text{det}\frac{N_T \Delta t}{4\pi L^2}\int_{E_{\gamma, \text{min}}}^{E_{\gamma, \text{max}}} \int_{E_{a,\text{min}}}^{E_{a,\text{max}}}
    \frac{\text{d}\Phi^\text{C}_a}{\text{d}E_a}\,\frac{\text{d}\sigma^\text{C}_\text{D}}{\text{d}E_\gamma}(E_{\gamma},E_a)\,\,\text{d}E_a \text{d}E_\gamma,  \\
   \cal{N}^\text{C}_\text{A}&=\int_{E_{a,1}}^{E_{a,2}} \frac{\text{d}\cal{N}^\text{C}_\text{A}}{\text{d}E_a} \, \text{d}E_a  \nonumber\\ 
  &=m_\text{det}\frac{N_T \Delta t}{4\pi L^2} \int_{E_{a,1}}^{E_{a,2}} 
  \frac{\text{d}\Phi^\text{C}_a}{\text{d}E_a}\,\sigma^\text{A}_\text{D}(E_{\gamma},E_a)\, \text{d}E_a, \\
  \cal{N}^\text{C}_\text{decay}&=\int_{E_{a,1}}^{E_{a,2}} \frac{\text{d}\cal{N}^\text{C}_\text{decay}}{\text{d}E_a} \, \text{d}E_a \nonumber\\
  &=  \frac{\mathcal{A}  \Delta t}{4\pi L^2} \int_{E_{a,1}}^{E_{a,2}} \frac{\text{d}\Phi^\text{C}_a}{\text{d}E_a}\,\mathcal{P}_\text{decay}\, \text{d}E_a,
\end{align}
for the detection mechanisms inverse Compton-like, axio-electric, and
$e^+ e^-$ decay, respectively. The integration limits $E_{a,1(2)}$ are
extracted from the ALP flux range, whereas the final integration over
$E_{\gamma}$ is performed using the limits
$E_{\gamma, \text{min}} =25$ keV---which we assume to be the minimum
detectable photon energy---and $E_{\gamma, \text{max}}$ that can be
obtained by inverting Eq.~(\ref{eq:Eaminmax}), leading to the simple
relation
$E_{\gamma, \text{max}} = y_\text{max}/(2 (E_{a,\text{max}}+m_e -|\vec
p_a|))$.
% -------------
% Sub-section
% -------------
\subsection{Production through nucleon-ALP coupling}
\label{sec:neutron-ALP-coupling}
A heavy nucleus disintegrates through $\alpha$ or $\beta$ decay or
fission processes, while the resulting daughter nucleus is typically left in
an excited state. It can then undergo fission, provided the state is above
the excitation energy for this process to take place, otherwise it
will in most cases decay by $\gamma$ emission. The study of nuclear
de-excitation follows from longitudinal ($L$), transverse ($T$),
transverse-longitudinal ($TL$) and transverse-transverse ($TT$)
nuclear responses. Their multipole expansion (projection in the
angular momentum basis) define the Coulomb multipole operator
$\widehat{M}_{JM}(q)$, the transverse electric and magnetic multipole
operators $\widehat{T}_{JM}^\text{el}(q)$ and
$\widehat{T}_{JM}^\text{mag}(q)$ and the longitudinal multipole
operator $\widehat{L}_{JM}(q)$. The angular momentum quantum numbers
$J$ and $M$ are determined by the angular momentum conservation
conditions $|J_f-J_i|\leq J \leq J_f+J_i$ and $M=M_f-M_i$, where $J_i$
and $J_f$ stand for the momenta of the initial and final states
involved in the nuclear excitation (or de-excitation) process (for
details see e.g.~\cite{Donnelly:1978tz,DeForest:1966ycn}).

For virtual photons --- as e.g. those involved in electron-nucleus scattering
processes --- all responses matter, while for real photons only
transverse electric and magnetic responses are relevant. These account
for electric (magnetic) transitions E$J$ (M$J$), with multi-polarities
determined by $J$. The lowest orders correspond to radiation emission
from charge ($J=0$, only electric), dipole ($J=1$) and quadrupole
($J=2$). In addition to the angular momentum conservation, transitions are
as well subject to parity conservation, for instance an E$J$ transition produces a
photon with parity $\Pi_\gamma=(-1)^J$, while a M$J$ transition a
photon with parity $\Pi_\gamma=(-1)^{J+1}$. Only photons for which
$\Pi_\gamma=\Pi_i\Pi_f$ is satisfied are allowed. If parity is not
changed, $\Delta \Pi=0$, only even (odd) electric (magnetic)
transitions are possible. If on the contrary $\Delta \Pi\neq 0$, only
odd (even) electric (magnetic) transitions are available.

\begin{figure}
  \centering
  \includegraphics[scale=0.85]{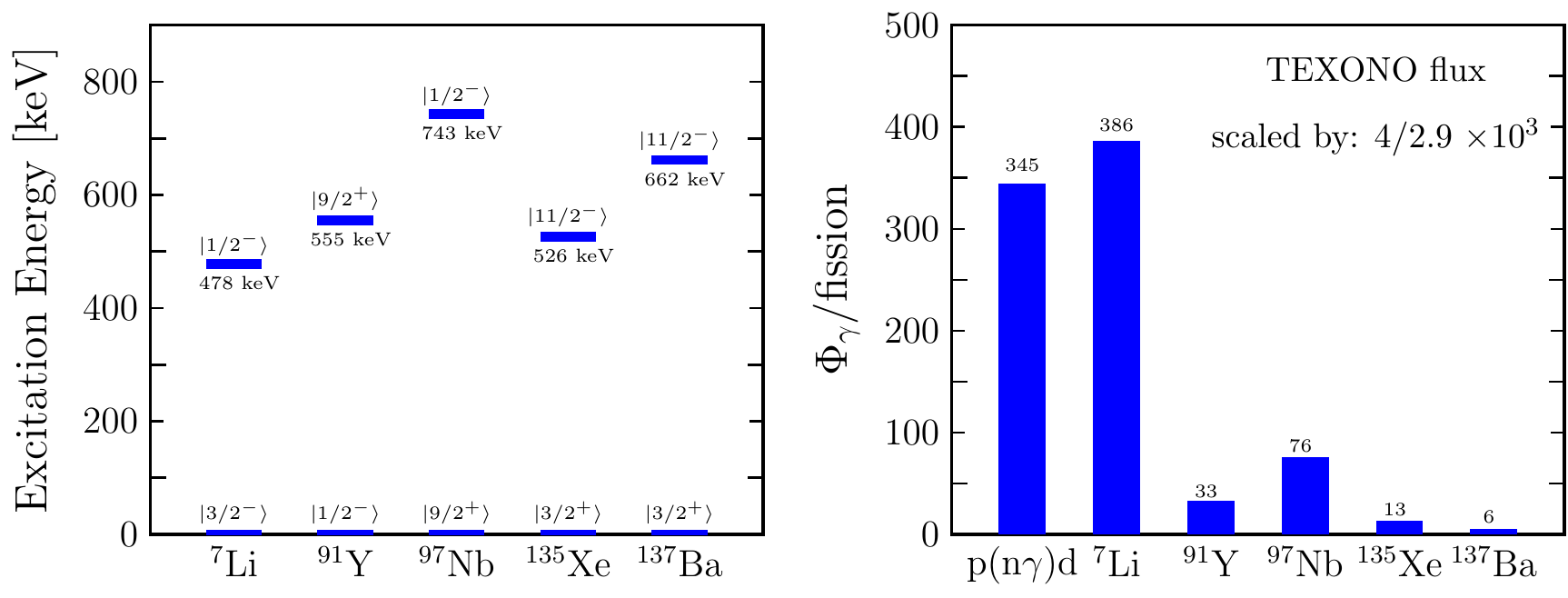}
  \caption{\textbf{Left graph}: $E=478\,$keV M$1$ transition from the
    excited state to the ground state of $^7$Li along with M$4$
    transitions from the excited to the ground state of $^{91}$Y,
    $^{97}$Nb, $^{135}$Xe and $^{137}$Ba (see text for more
    details). These transitions involve $\gamma/a$ lines at
    $555\,$keV, $743\,$keV, $526\,$keV and $662\,$keV,
    respectively. \textbf{Right graph}: Photon flux for a 4 GW power
    plant and for each of the channels considered in our
    analysis. They follow from the TEXONO flux rescaled by 4/2.9~\cite{Chang:2006ug}.}
  \label{fig:magnetic-transitions}
\end{figure}

If ALPs couple to nucleons, nuclear de-excitation can proceed as well
through ALP emission. Since ALPs are pseudoscalars
($|J^\Pi\rangle =|0^-\rangle$) they can be emitted only in magnetic
transitions~\cite{Weinberg:1977ma}, in contrast to photons which are
produced in electric transitions as well. The possible quantum numbers
they carry away are therefore $|0^-\rangle$, $|1^+\rangle$,
$|2^-\rangle$, $|3^+\rangle,...$~\cite{Donnelly:1978ty}. To the best of our knowledge, searches for axions in magnetic transitions were pioneered by
Zehnder in a nuclear decay experiment which used the $664\,$keV M$4$
transition from the $|11/2^-\rangle$ excited state to the
$|3/2^+\rangle$ ground state of $^{127}\text{Ba}^*$~\cite{Zehnder:1981qn}. Subsequent searches were conducted at the Bugey
reactor ($\sim 2.8\,$GW) using the neutron capture M$1$ isovector
transition along with M$4$ transitions in $^{97}\text{Nb}$,
$^{91}\text{Y}$, $^{137}\text{Ba}$ and $^{135}\text{Xe}$ (see left
graph in Fig.~\ref{fig:magnetic-transitions})~\cite{Cavaignac:1982ek}. Searches focusing on the 1115 keV M1 transition
from the $|5/2^-\rangle$ excited state to the $|3/2^-\rangle$ ground
state of $^{65}$Cu were performed as well~\cite{Lehmann:1982bp}, while the same approach was later followed by Avignone
et. al.~\cite{Avignone:1986vm}. More recently, similar searches were pursued at
the Kuo-Sheng Nuclear Power Station by the TEXONO collaboration, which
in addition to the transitions used at Bugey, considered a M$1$ transition in
$^7\text{Li}$ (see left graph in Fig.~\ref{fig:magnetic-transitions})~\cite{Chang:2006ug}.

In our analysis we employ the fluxes reported by TEXONO: below we
discuss in more detail the magnetic transitions from which they are generated. Thermal neutron capture on proton in the cooling
water, $\mathrm{p+n\to d+\gamma}$, produces a monochromatic line at $2230\,$keV
(deuteron binding energy). The deuteron ground state has magnetic
dipole and electric quadrupole moments, the emitted $\gamma$ is
therefore mainly M1. On the other hand, the $^7\text{Li}^*$ de-excitation involves M1+E2
transitions from the $|1/2^-\rangle$ first excited state to the
$|3/2^-\rangle$ ground state, and so again the transition is mainly
M1. For $^{91}\text{Y}^*$ and $^{97}\text{Nb}^*$ de-excitations from
the first excited state to the ground state, as shown in
Fig.~\ref{fig:magnetic-transitions}, correspond to M4
transitions. Finally, $^{135}\text{Xe}^*$ and $^{137}\text{Ba}^*$
de-excitations involve M4 transitions from the second excited state to
the ground state. Note that for $^{135}\text{Xe}$, de-excitation from
the $|1/2^+\rangle$ first excited state to the ground state
corresponds to M1+E2 transitions. If it was available, the flux from
this transition (mainly M1) would be much larger than that from the M4
transition. The corresponding fluxes are shown in the right graph of
Fig.~\ref{fig:magnetic-transitions}.

ALP production depends on the ALP-nucleon couplings and on the ALP
mass, while the relevant ALP flux is determined through the competition of
the axion-to-photon emission rates, as
first calculated in~\cite{Donnelly:1978ty}. In contrast to M4
transitions, neutron capture isovector M1 transitions (pn$\to$
d$\gamma$) depend only on kinematics and the vector ALP-nucleon
coupling~\cite{Barroso:1981bp}
\begin{equation}
  \label{eq:m1-transition-neutron-capture}
  \left(\frac{\Gamma_a}{\Gamma_\gamma}\right)_{\text{pn}}
  =\frac{1}{2\pi\alpha}\left(\frac{|\vec p_a|}{|\vec p_\gamma|}\right)^3
  \left(\frac{g_{ann}^{(1)}}{\mu_1}\right)^2\ ,
\end{equation}
where $\mu_1$ refers to the isovector magnetic moment which follows
from the neutron and proton magnetic moments, $\mu_1=\mu_\text{p}-\mu_\text{n}=4.71$
in nuclear magneton units $\mu_N=1/2m_\text{p}$ ($m_\text{p}$ is the proton mass)~\cite{Avignone:1986vm}. For M$J$ transitions this ratio instead
involves nuclear physics input, and in the long-wavelength limit it reads~\cite{Chang:2006ug}
\begin{equation}
  \label{eq:axion-to-photon-rate}
  \left(\frac{\Gamma_a}{\Gamma_\gamma}\right)_{\text{M}J}
  =\frac{1}{\pi\alpha}\left(\frac{1}{1+\delta^2}\right)
  \left(\frac{J}{J+1}\right)
  \left(\frac{|\vec p_a|}{|\vec p_\gamma|}\right)^{2J+1}
  \left(\frac{g_{ann}^{(0)} \kappa  + g_{ann}^{(1)}}
    {(\mu_0-1/2)\kappa + (\mu_1+\eta)}\right)^2\ .
\end{equation}
Here, $\mu_0$ refers to the isosinglet magnetic moment normalized to
$\mu_N$ and obtained as $\mu_0=\mu_\text{p}+\mu_\text{n}=0.88$~\cite{Avignone:1986vm}. The remaining $\delta$, $\eta$ and $\kappa$ 
parameters are nuclear-structure dependent. In particular, the multipole mixing ratio
$\delta$ is determined by the E$J^\prime$-to-M$J$ ($J^\prime=J+1$)
transition probabilities ratio, while $\kappa$ and $\eta$ are defined by
ratios of the orbital ($\widehat L$) and spin ($\widehat\sigma$)
nuclear operators reduced matrix elements, namely~\cite{Avignone:1986vm}
\begin{align}
  \label{sec:eta-beta-parameter}
   \eta= - \frac{\langle J_f||\sum_{k=1}^A \widehat L(k)\tau_3(k)||J_i\rangle}
  {\langle J_f||\sum_{k=1}^A\widehat \sigma(k)\tau_3(k)||J_i\rangle}\ ,
  \qquad
  \kappa=\frac{\langle J_f||\sum_{k=1}^A \widehat \sigma(k)||J_i\rangle}
  {\langle J_f||\sum_{k=1}^A\widehat \sigma(k)\tau_3(k)||J_i\rangle}\ ,
\end{align}
where $J_{i,f}$ refer to initial and final state angular momenta and
the sum runs over the total number of nucleons.

Values of $\delta$, $\eta$ and $\kappa$ for the $^7\text{Li}^*$ M1
transition have been calculated in Refs.~\cite{Krcmar:2001si} and also adopted
by the TEXONO collaboration in their analysis in~\cite{Chang:2006ug}. The transition is an admixture of M1+E2, for
which the transition probability ratio is small $\delta\ll 1$ and therefore it
can be neglected, while the estimated values for $\eta$ and $\kappa$ are
0.5 and 1.0. For the remaining transitions $\delta=0$, as expected
since they are purely magnetic, while values for $\eta$
and $\kappa$ are not available.  However, using results for $^{65}$Cu,
$^{57}$Fe, $^{55}$Mn and $^{23}$Na---calculated in
Refs.~\cite{Avignone:1986vm,Haxton:1991pu}---TEXONO estimated these
values to be $\eta=1$ and $\kappa=-3$ for $^{91}$Y and $^{97}$Nb and
$\eta=-1$ and $\kappa=1$ for $^{135}$Xe and $^{137}$Ba~\cite{Chang:2006ug}. We adopt these values in our analysis.

With the values of the nuclear-structure-dependent parameters
specified, we can now write the ALP flux for the $i-$th transition 
\begin{equation}
  \label{eq:axion-line}
  \left(\frac{\text{d}\Phi^\text{MT}_a}{\text{d}E_a}\right)_i=\phi_a^i\,\delta(E_{\gamma^{\prime}}-E_a)=R_f \Phi_\gamma^i
  \left(\frac{\Gamma_a}{\Gamma_\gamma}\right)_i \mathcal{P}_\text{surv} \,  \delta(E_{\gamma^{\prime}}-E_a) 
  \qquad (i=\text{p}(\text{n},\gamma)\text{d},~\text{M}J)\ ,
\end{equation}
with $\Phi_\gamma^i$ the photon flux per fission for the $i-$th
transition specified in the right graph in
Fig.~\ref{fig:magnetic-transitions} and $R_f$ the fission rate, which
we assume that proceeds entirely from $^{235}$U and estimate to be
$1.248\times 10^{20}\;$fissions/s for a 4 GW nuclear reactor. 
The ALP yield at the detector depends on the detection mechanism
(see Tab.~\ref{tab:processes-and-couplings}). In the presence of only
$g_{ann}^{(a)}$ couplings, the  detection would be possible only through ALP
nuclear excitation. This will require the detector to be built from
one of the daughter isotopes from which the ALP was produced, to
enable ALP absorption. Since this does not seem to be a viable
possibility, in this case we abandon our minimal assumption analysis
and instead we rely on detection through Primakoff, Compton-like,
axio-electric processes and decays. The nuclear de-excitation analysis is
therefore sensitive to coupling pairs and the ALP mass, in contrast to the
Primakoff and Compton-like analyses which determine
sensitivities on a single coupling and the ALP mass. 

The ALP detection rate for the $i-$th transition proceeds as always
from a convolution of the ALP flux and the differential cross section
of the corresponding detection process. For detection through inverse
Primakoff, Eq.~(\ref{eq:diff-event-rate-scattering-Primakoff}) applies
and so we write
\begin{equation}
  \label{eq:differential-rate-transitions}
  \left(\frac{\text{d}\cal{N}_\text{P}^\text{MT}}{\text{d}E_a}\right)_i=m_\text{det}
  \frac{N_T \Delta t}{4\pi L^2}\int 
  \left(\frac{\text{d}\Phi^\text{MT}_a}{\text{d}E_a}\right)_i
  \,\frac{\text{d}\sigma^\text{P}_\text{D}}{\text{d}E_\gamma} \text{d}E_\gamma\ .
\end{equation}
Using the ALP flux in Eq.~(\ref{eq:axion-line}) the ALP yield then
becomes
\begin{equation}
  \label{eq:ALP-yield-MJ-T-Primakoff}
  ({\cal N}_\text{P}^\text{MT})_i=m_\text{det}\,\frac{N_T \Delta t}{4\pi L^2}\,\phi_a^i\,
  \sigma_\text{D}^\text{P}\,\ .
\end{equation}
The same applies as well for detection through inverse Compton-like
and axio-electric processes, for which we then write
\begin{equation}
  \label{eq:compton-like-axio-electric-MJ-transitions}
    ({\cal N}_\text{C}^\text{MT})_i=m_\text{det}\,\frac{N_T \Delta t}{4\pi L^2}\,\phi_a^i\,
     \sigma^\text{C}_\text{D} \ ,
    \qquad
    ({\cal N}_\text{A}^\text{MT})_i=m_\text{det}\,\frac{N_T \Delta t}{4\pi L^2}\,\phi_a^i\,
    \sigma^\text{A}_\text{D} \ ,
\end{equation}
and to ALPs detected through their decays into diphoton or
electron-positron pairs:
\begin{equation}
  \label{eq:compton-like-axio-electric-MJ-transitions_decays}
    ({\cal N}_\text{decay}^\text{MT})_i=\frac{\mathcal{A}  \Delta t}{4\pi L^2}\,\phi_a^i\,
    \mathcal{P}_\text{decay} \ .
\end{equation}

% ------------
% Section III
% ------------
\begin{table}
  \centering
  \begin{tabular}{|c|c|c|c|c|c|c|}\hline
    \textbf{P[GW]}&\textbf{PM}&\textbf{TM}
    &$\boldsymbol{m_\text{\bf det}} \text{\bf [kg]}$
    &$\boldsymbol{L} \text{\bf[m]}$
    &$\boldsymbol{L_\text{\bf det}} \text{\bf [cm]}$
    &\textbf{bkg [1/keV/day/kg]}\\\hline\hline
    4 & $^{235}$U & Ge & 10 & 10 & 50 & 10-100\\\hline
    8 & $^{235}$U & Xe & 10$^3$ & 10 & 140 & 1-10\\\hline
  \end{tabular}
  \caption{Production and detection parameters used in our analysis. 
    Upper values refer to current (or near-future) experiments, while
    those in the lower row to a next-generation experiment. 
    These values---used for definitiveness---should be understood 
    as representative of the actual experimental parameters shown 
    in Tabs. \ref{tab:reac_exps_prod} and \ref{tab:reac_exps_det}. 
    \textbf{P} refers to power, \textbf{PM} to 
    production material and \textbf{TM} to target material. Our 
    analysis includes two background hypotheses and we assume the
    detector to be $100\%$ efficient.}
  \label{tab:our-parameters}
\end{table}
\section{Experimental sensitivities}
\label{sec:discovery-reach}
In this section we discuss sensitivities that the experiments listed
in Tab.~\ref{tab:reac_exps_prod} and \ref{tab:reac_exps_det} could in
principle achieve. Rather than focusing on a particular case we adopt
a generic analysis whose results we believe provide a general
picture. To do so we fix the relevant experimental production and
detection parameters as shown in Tab.~\ref{tab:our-parameters}. Upper
row values provide sensitivities for ongoing or near-future
experiments, while lower row values represent what could be achieved in %what we
%regard as 
a next-generation experiment.  Given an ALP coupling, our
analysis includes the processes shown in
Tab.~\ref{tab:processes-and-couplings}, the exception being the case
of ALP production through nuclear de-excitations for the reasons we
have discussed in Sec. \ref{sec:neutron-ALP-coupling}. Calculation of
sensitivities is done with the aid of the following $\chi^2$ parameter
function
\begin{equation}
  \label{eq:chiSq}
  \chi^2=\frac{\mathcal{N}_\text{ALP}^2}{\cal{N}_\text{ALP} + \cal{N}_\text{bkg}}. 
\end{equation}
 Here by $\cal{N}_\text{ALP}$ we generically refer to the ALP event yield, obtained through any of the
processes that we have discussed in Sec. \ref{sec:mechanisms}, while $\cal{N}_\text{bkg}$  refers to the background
hypothesis adopted (see Tab.~\ref{tab:our-parameters}).

\begin{figure}
  \centering
  \includegraphics[width = 0.7 \textwidth]{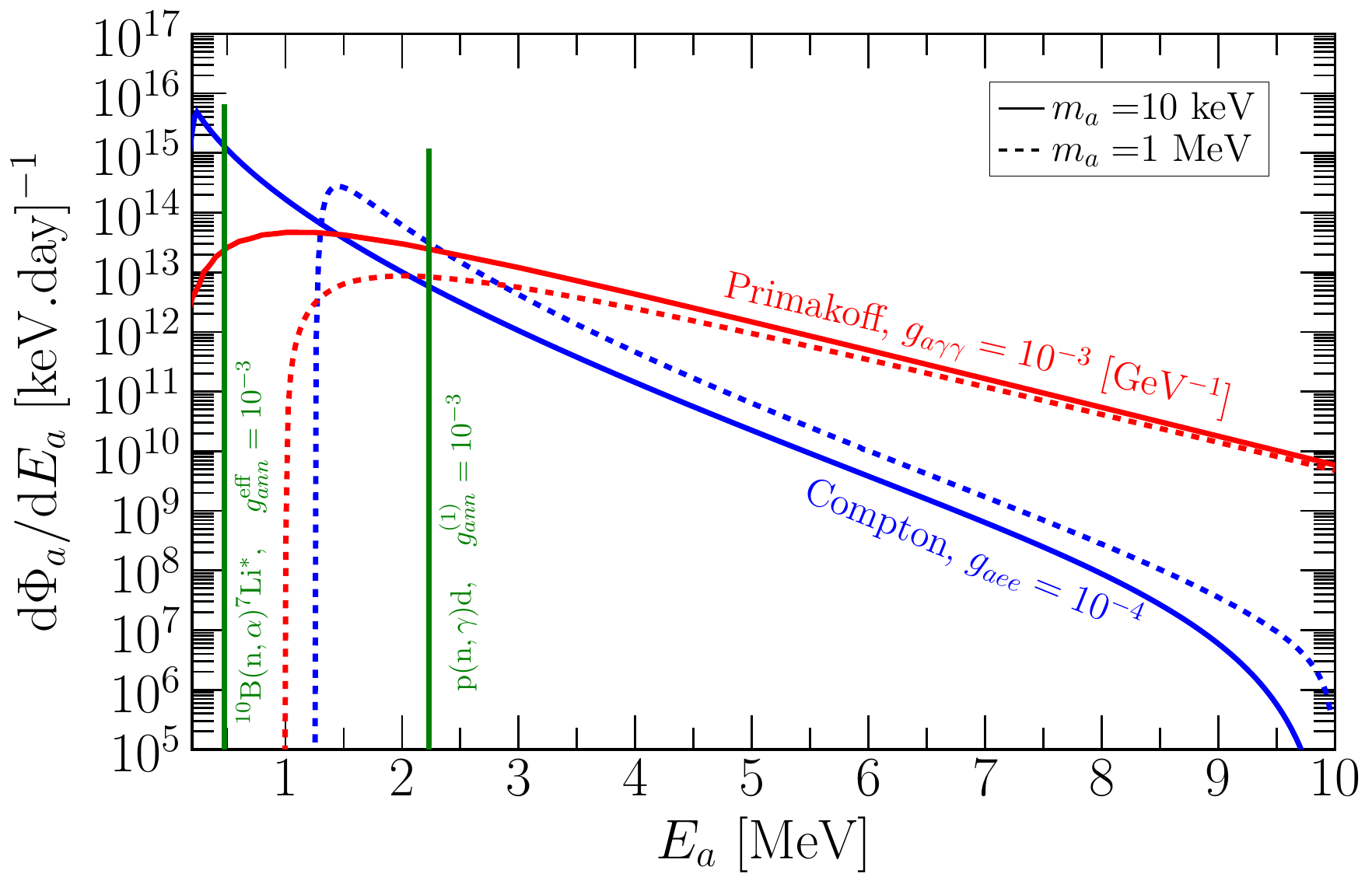}
  \caption{Typical ALP flux generated in a 4 GW nuclear power plant
    through Primakoff and Compton-like scattering processes as well as
    through neutron capture in proton and $^7$Li$^*$ de-excitation
    (nuclear dominant M1 transition processes). Here we have assumed
    $L=10\,$m (reactor-detector distance), which is the value that we have
    used for the calculation of sensitivities in both current and future scenarios.}
  \label{fig:ALP-flux}
\end{figure}

\begin{figure}
  \centering
  \includegraphics[width=0.75\textwidth]{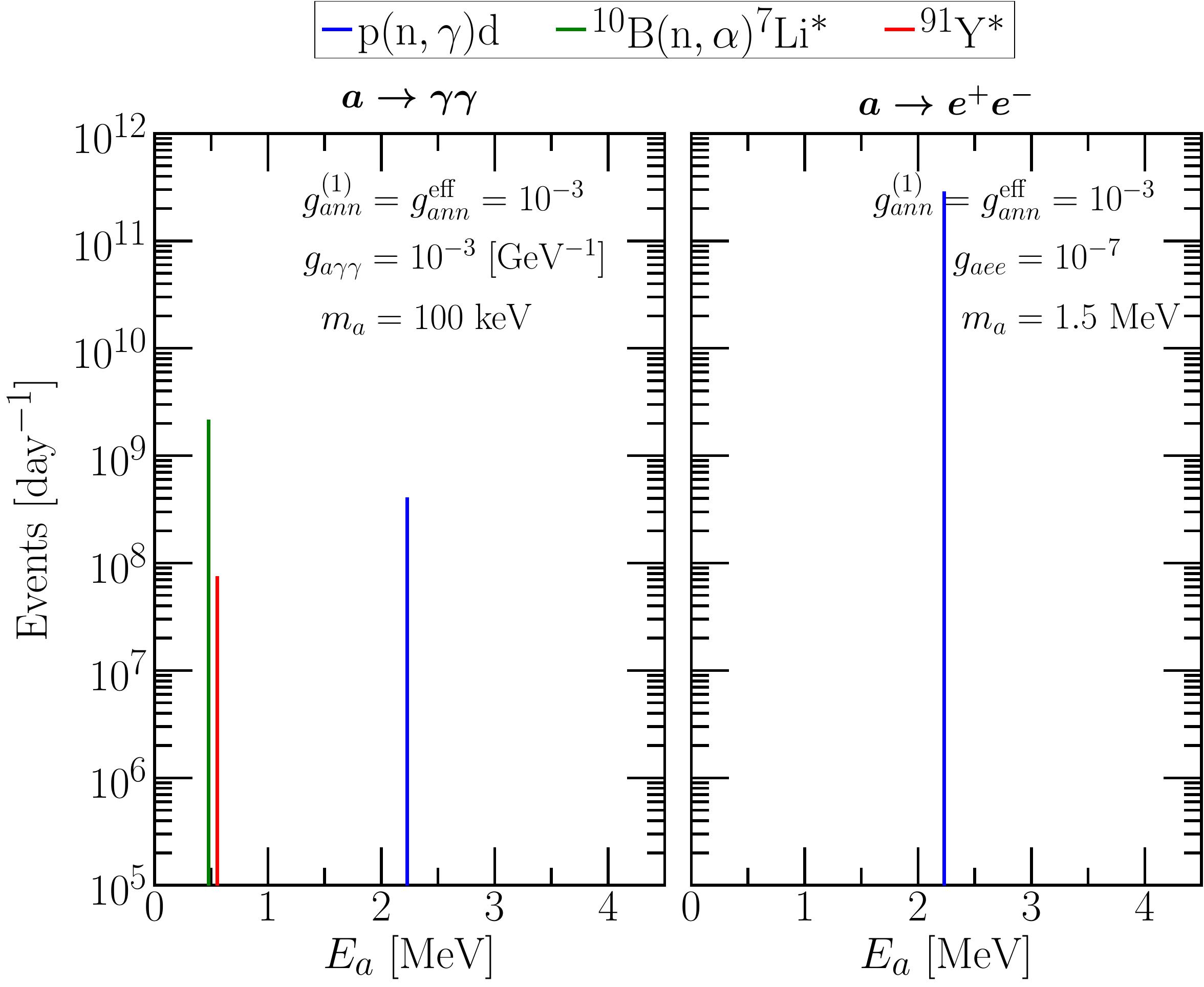}
  \includegraphics[width=\textwidth]{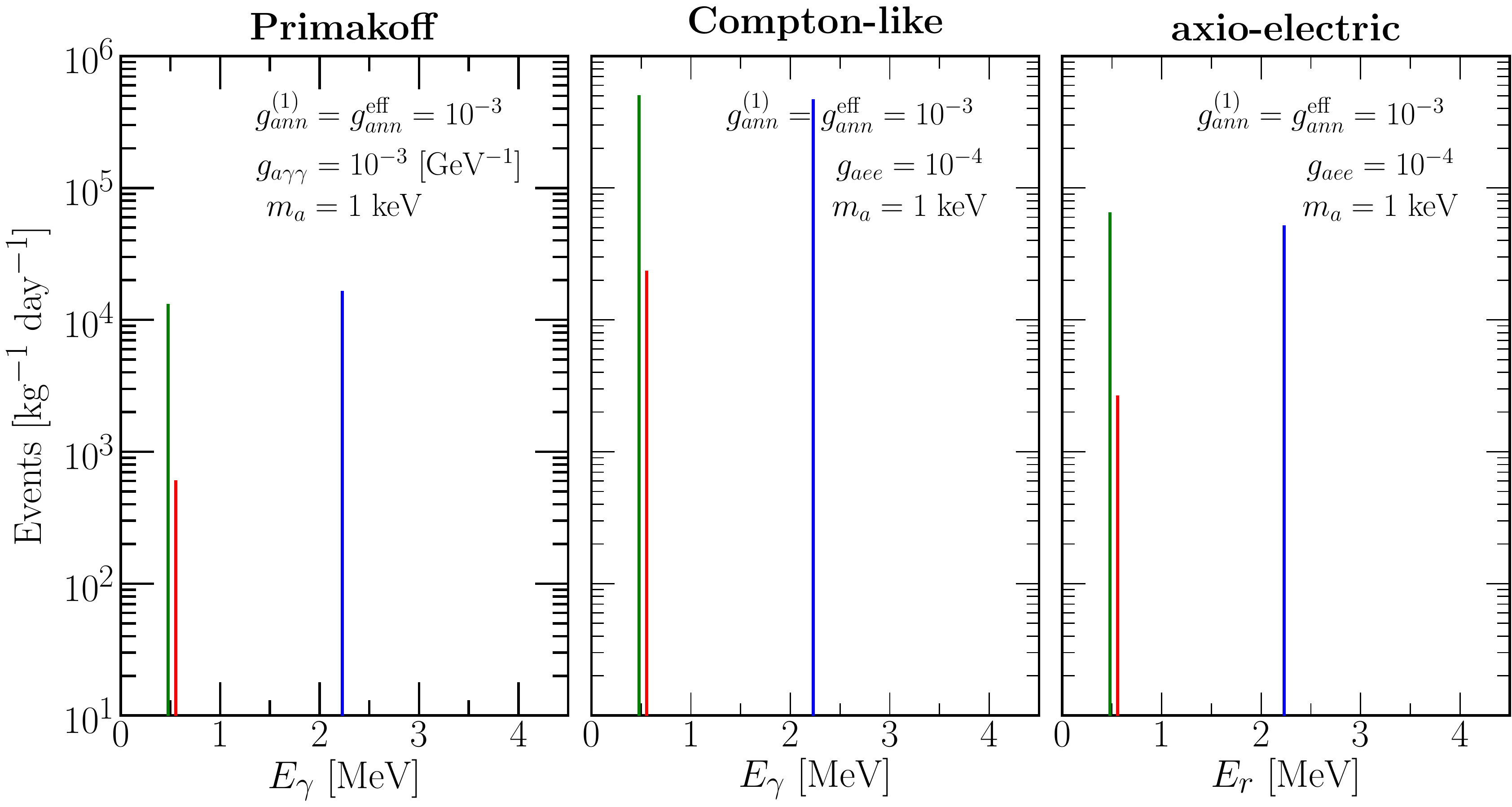}
  \caption{Event yields for ALP production through magnetic
    transitions and detection through: decays $a\to \gamma\gamma$ and
    $a\to e^+e^-$ (left and right upper graphs), inverse-Primakoff
    scattering $a+N\to \gamma+N$ (left lower graph), Compton-like
    scattering $a+e\to \gamma+e$ (middle lower graph) and
    axio-electric absorption $a+e+Z\to \gamma+e+Z$ (right lower
    graph). The event yield has been calculated for a 4 GW reactor
    power plant assuming $L=10\,$m. Here, $g_{ann}^\text{eff}=g_{ann}^{(0)} \kappa  + g_{ann}^{(1)}$.}
  \label{fig:magnetic-transitions-event-yield-example}
\end{figure}

We begin our analysis by comparing the size of the ALP flux generated
by Primakoff and Compton-like scattering as well as by M1 transitions
(neutron capture in proton and $^{7}$Li$^*$ de-excitation). Figure
\ref{fig:ALP-flux} shows the result for parameter choices as displayed
in the plot. One can see that a large amount of ALPs can be produced
through any of the mechanisms we consider.  Despite being parameter
dependent, the size of the flux can become large in certain regions of
the parameter space, as shown in Fig.~\ref{fig:ALP-flux}.  Indeed,
reactor-based neutrino experiments have the potential of reaching
large sensitivities, further motivating our present study. Note that
M4 transitions produce a rather comparable flux to that from M1
de-excitation processes. However we found that their event yield is
small (see Fig.~\ref{fig:magnetic-transitions-event-yield-example}),
and so in the ALP-nucleon coupling analysis we consider only
$\mathrm{p(n,d)\gamma}$.

\begin{figure}[t!]
  \centering
  \includegraphics[width=0.49 \textwidth]{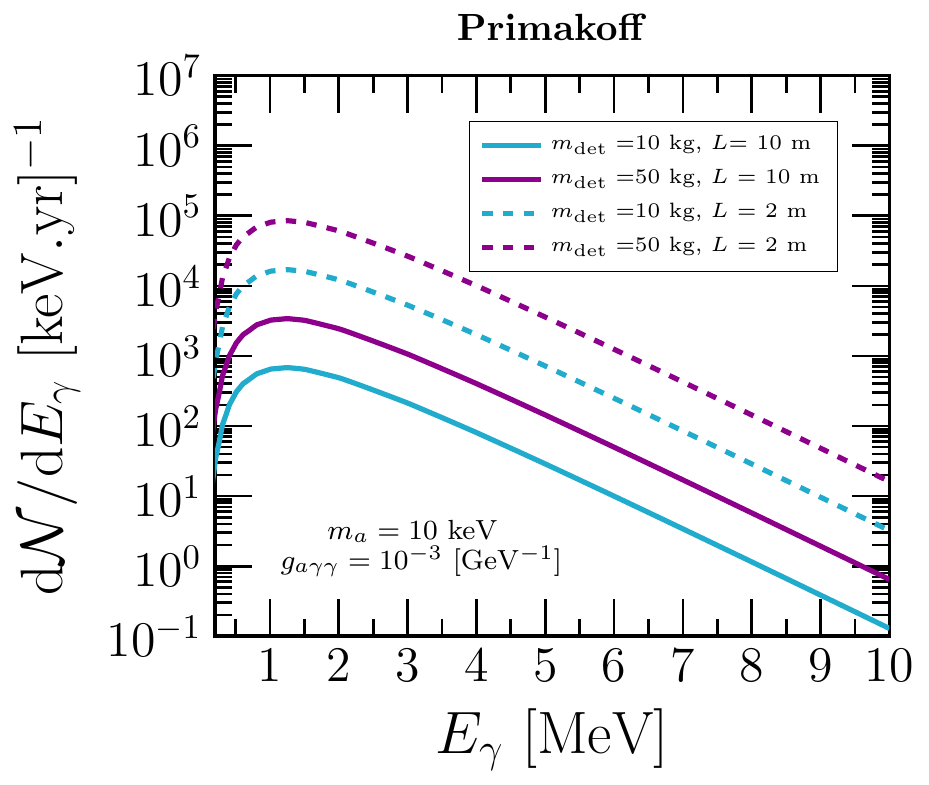}
  \includegraphics[width=0.49 \textwidth]{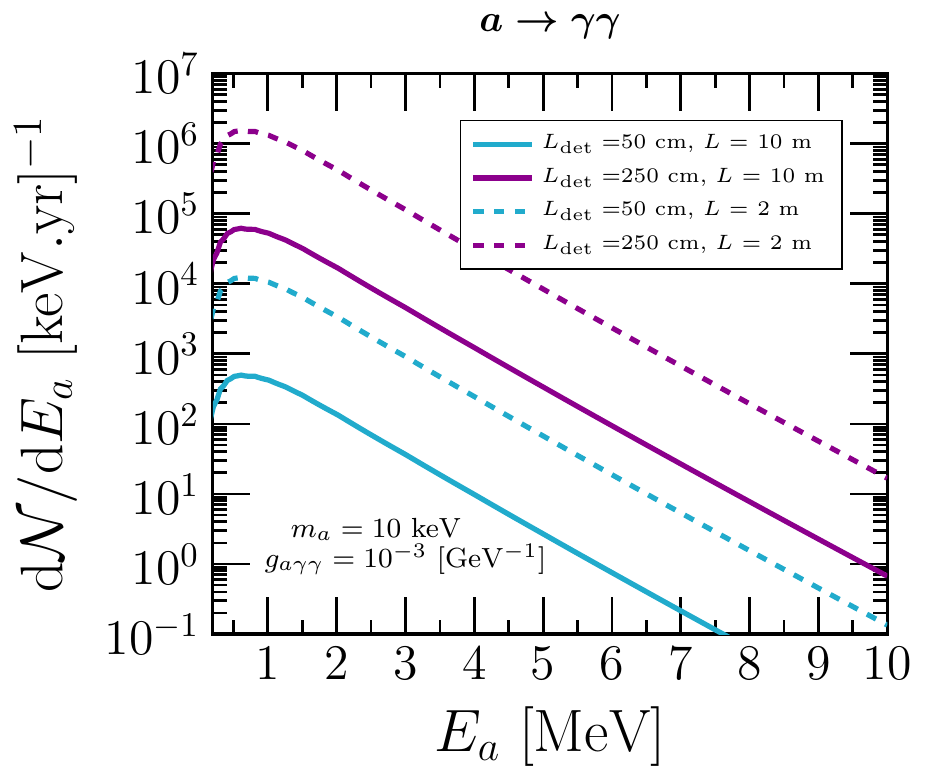}
  \caption{\textbf{Left graph}: ALP differential event rate for the
    case of production and detection through Primakoff scattering
    (applicable to ALP models with dominant coupling  only to photons)
    for different values of the reactor-detector baseline and detector
    masses. \textbf{Right graph}: Same as in left graph, but for
    detection through ALP decay, $a\to \gamma\gamma$. The
    reactor-detector baseline as well as the detector length
    $L_\text{det}$ vary in this case.}
  \label{fig:DER-Primakoff}
\end{figure}

With large ALP fluxes, a large ALP event yield crucially depends on
whether they can reach and decay within the detector -- which is
accounted for by the probabilities in Eq.~(\ref{eq:survival}). While a number of
ALPs can as well interact with the detector shielding thus
diminishing the expected ALP flux, for the sake of simplicity we neglect those
interactions and we assume the interaction probability to be suppressed
($\mathcal{P}_\text{det}\to 0$). In order to demonstrate the
importance of the different detector variables, we have calculated
differential event rates for the Primakoff and decay
$a\to \gamma\gamma$ cases (representative of models with dominant
$g_{a\gamma\gamma}$ coupling\footnote{A tree level
  $g_{a\gamma\gamma}$ coupling generates through one-loop vertex
  diagrams $g_{aee}^\text{1-loop}$ and $g_{app}^\text{1-loop}$
  couplings. These couplings being controlled by $g_{a\gamma\gamma}$
  are at least suppressed by the $1/(16\pi^2)$ loop suppression
  factor. The same argument applies for models with tree level
  $g_{aee}$ and $g_{ann}$ couplings. The other ALP couplings will
  arise at the radiative level, and so their possible effects will be
  suppressed.}) varying the reactor-detector distance $L$, the detector mass
$m_\text{det}$ and the detector length $L_\text{det}$. Results are shown
in Fig.~\ref{fig:DER-Primakoff}.

One can see that the reactor-detector distance plays an important role
regardless of the process, as expected given the quadratic flux
dependence on that variable. Moving from $10\,\text{m}$ to
$2\,\text{m}$ enhances the event rate by a factor 25, due to the usual isotropic reduction factor $1/(4\pi L^2)$. For
$a+N\to \gamma+N$ scattering, the detector mass plays an important role as
well although its effect is less pronounced than the reactor-detector
distance. For $a\to \gamma\gamma$ decay the
detector length is instead a key variable. As shown in the right graph
in Fig.~\ref{fig:DER-Primakoff} a factor 5 increase in $L_\text{det}$
has a large impact in the event rate, resulting in an enhancement of
about two orders of magnitude. From these results one can fairly
conclude that, if backgrounds are well understood and under control,
ideally a short reactor-detector distance combined with a large
detector mass and length would maximize the ALP event rate.

\begin{figure}
  \centering
    \includegraphics[width =  \textwidth]{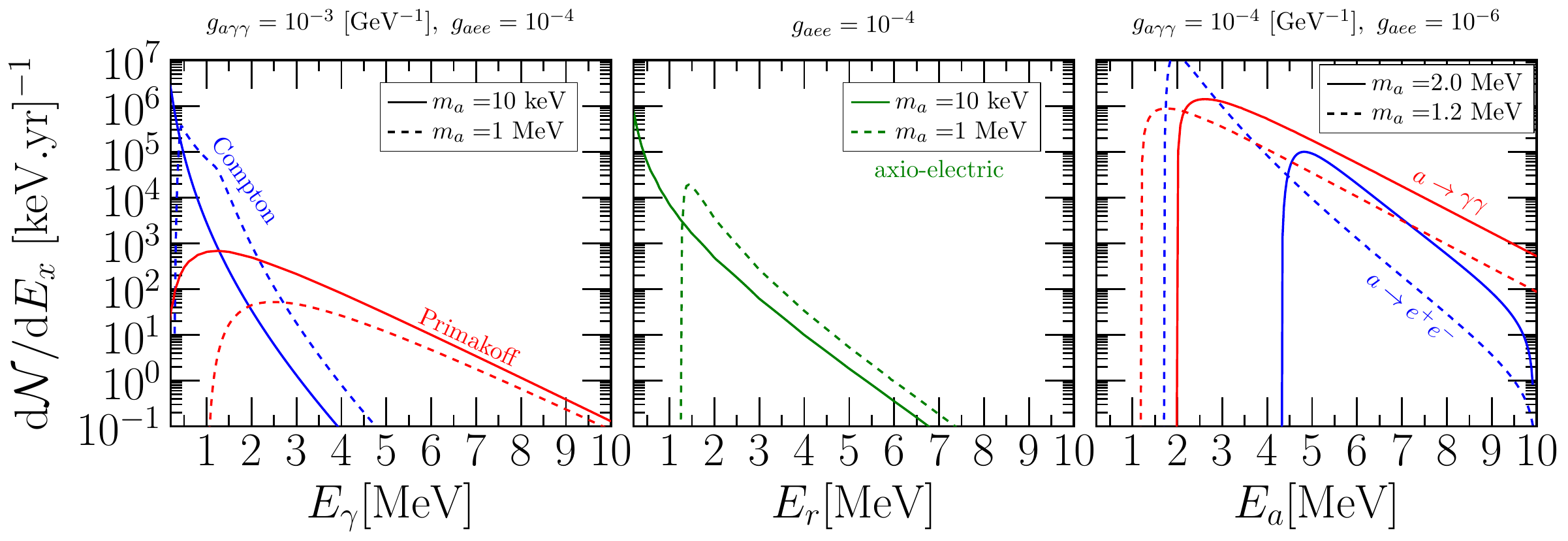}
    \caption{Differential event rate for Compton-like and Primakoff
      scattering (left graph), axio-electric absorption (middle graph)
      and ALP decay to electron-positron and photon pairs (right graph).}
  \label{fig:gaee-processes-example}
\end{figure}

As we have already stressed, for scenarios with dominant $g_{aee}$
coupling three possible detection mechanisms are possible:
Compton-like scattering, axio-electric absorption and ALP decay to
electron-positron pairs. To compare their relative importance we have calculated
their differential event rates for particular parameter choices.
Results are shown in Fig.~\ref{fig:gaee-processes-example} plotted
versus the relevant kinematic variable in each case. Compton-like and
axio-electric absorption have about the same energy behavior, although
in the former case the relevant variable is the photon energy while in the latter is the electron
recoil energy. As it will become clear when discussing $g_{aee}$ and
$g_{aee}\times g_{ann}^{(1)}$ sensitivities, due to their close energy
behavior they contribute similarly to the event
yield. A comparison of these processes with ALP decay to electron-positron pairs
shows that at large energies and for large ALP masses the decay
dominates over the absorption and scattering processes. Inverse
Primakoff compared with Compton-like scattering extends all over the
photon energy range, but produces less events. This results in
ALP-electron sensitivities being slightly better. The same applies for
decays.

For scenarios with dominant $g_{ann}$  coupling the event yield depends
on the detection mechanism. For illustration, we have calculated for
particular parameter space point values the different yields. Results
are displayed in
Fig.~\ref{fig:magnetic-transitions-event-yield-example}. First of all
one can see the difference between M1 and M4 transitions. In all cases
the event yields from neutron capture in proton and $^7$Li$^*$
de-excitation exceeds that from $^{91}$Y$^*$ by more than one order of
magnitude, a trend that applies for the other M4 transitions as
well.

Although the ALP event rate from decay to final state photons has no
kinematic constraint, its size does depend on the ALP mass as expected
from Eq.~(\ref{eq:decay-widths}). The absence of a kinematic limit
allows all transitions to generate such a process, with a monochromatic event yield characterized by the energy 
of the corresponding
transition.  Decay to electron-positron pairs is instead subject to the kinematic
constraint $m_a>2m_e$, and so only $\mathrm{p(n,\gamma)d}$ can induce such
process. Although involving different numerical values, comparison of
the left upper graph with the lower left graph in Fig.~\ref{fig:magnetic-transitions-event-yield-example} allows to see that
$a\to \gamma\gamma$ can play an important role, actually dominant once
$m_a\gtrsim 10^4\,$eV (this will become apparent when discussing
$g_{a\gamma\gamma}$ experimental sensitivities). Comparison of the
right upper graph and the middle and right lower graphs enables
establishing the same conclusion, decay to electron pair dominates as
soon as the channel is kinematically open. Comparison of the middle
and right lower graphs in turn demonstrates that Compton-like
scattering plays a more important role than axio-electric absorption
does, with the event rate exceeding about an order of magnitude.

\begin{figure}
  \centering
  \includegraphics[scale=0.65]{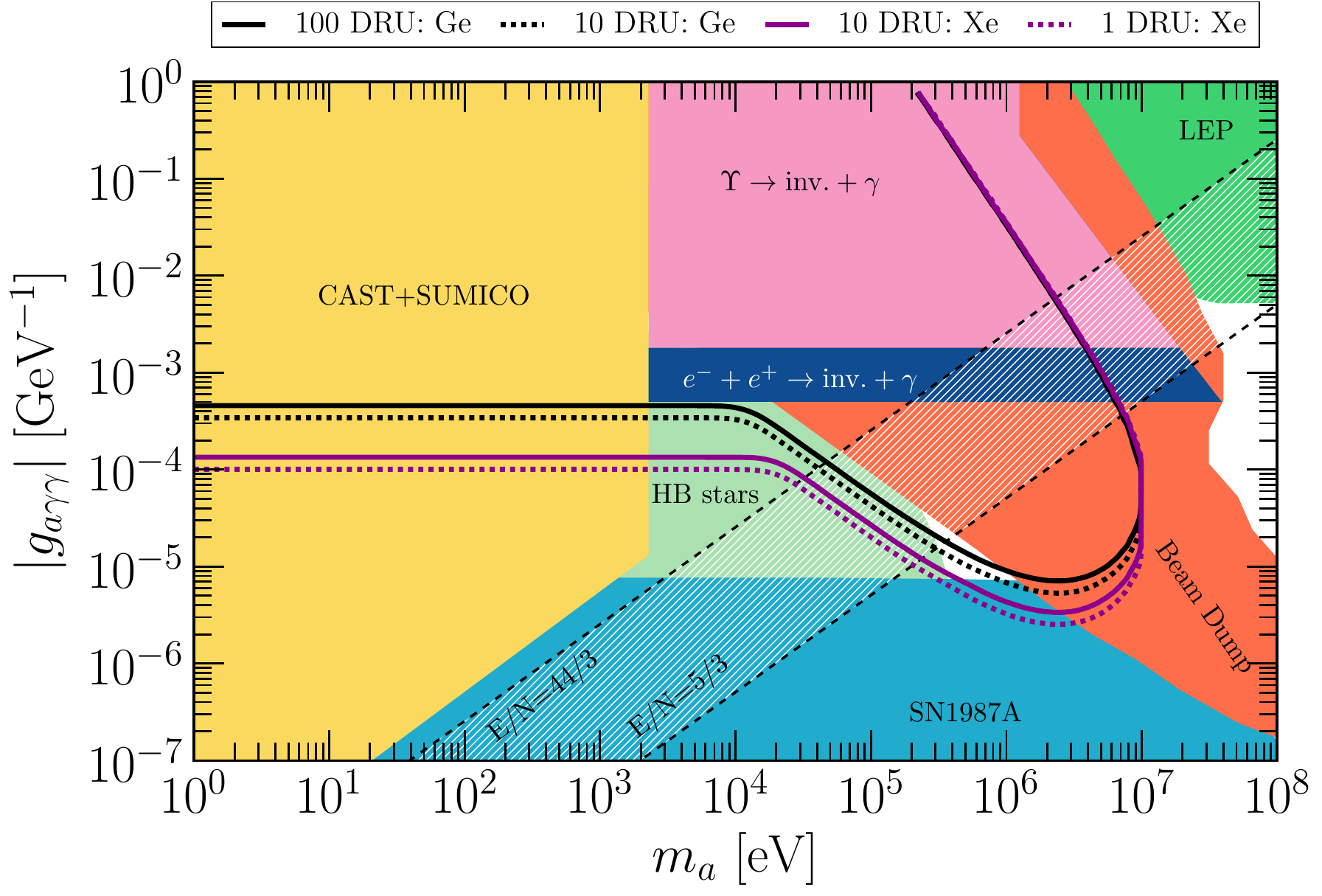}
  \caption{Photon-ALP coupling $90\%$CL sensitivities along with
    relevant laboratory, astrophysical and cosmological limits that
    apply in the region of interest, adapted from
    Ref.~\cite{Jaeckel:2015jla,Carenza:2020zil,Lucente:2020whw}.  Black contours correspond to
    $90\%$CL sensitivities achievable with ongoing or near-future
    experiments (see Tab.~\ref{tab:reac_exps_det}). Purple contours
    instead indicate $90\%$CL sensitivites that could be achieved in a
    next generation experiment. Solid (dotted) contours refer
    to sensitivities under the assumption of a background rate of 100 DRU (10 DRU) for
    ongoing or near-future experiments, while to 10 DRU (1 DRU) for a
    next generation setup. See the text for more details. For
    completeness we show as well the region covered by QCD axion
    models, with QCD and electromagnetic anomaly coefficients fixed
    according to Ref.~\cite{DiLuzio:2016sbl}.}
  \label{fig:gagg-sensitivities}
\end{figure}

We now turn to sensitivities that could be achieved in models with
dominant $g_{a\gamma\gamma}$ coupling. With the production and
detection specifications we have assumed, the region of interest that
can be explored corresponds to
$g_{a\gamma\gamma}\gtrsim 10^{-6}\,\text{GeV}^{-1}$ and
$m_a\subset [1,10^7]\,\text{eV}$. In this region there are various
laboratory as well as astrophysical and cosmological
constraints that apply. Relevant laboratory limits in the ``high'' mass region
follow from beam-dump experiments, invisible $\Upsilon$ decays and
$e^+e^-\to \text{inv}+\gamma$ searches~\cite{Krasny:1987eb,Dobrich:2017gcm,Bjorken:1988as,
  Antreasyan:1990cf,Aubert:2008as,Hearty:1989pq}. In the ``low'' mass
region instead limits from CAST and SUMICO are the most important~\cite{Arik:2008mq,Minowa:1998sj}. Laboratory limits shown in
Fig.~\ref{fig:gagg-sensitivities} are adapted from
Ref.~\cite{Jaeckel:2015jla}. Although our analysis applies for ALPs,
for completeness we show the region of hadronic QCD axion models
defined by $\text{E/N}\subset [5/3,44/3]$~\cite{DiLuzio:2016sbl}.

Constraints from astrophysical sources are derived from SN1987A and
horizontal branch (HB)
stars~\cite{Brockway:1996yr,Raffelt:1985nk,Raffelt:1987yu,Ayala:2014pea,Carenza:2020zil,Lucente:2020whw}. SN1987A
limits follow from SN energy loss, which in the absence of ALPs is
driven by neutrino emission, and from the visible signal that ALP
emission and subsequent decay to photons will generate. Given the
temperature of the SN environment, these bounds extend up to ALP
masses of the order of $10^8\,$eV. Horizontal branch stars limits
follow, instead, from stellar cooling arguments. The presence of the
new coupling opens a channel that accelerates the consumption of
helium resulting in a reduction of the HB stars
lifetime~\cite{Raffelt:1996wa}. It is worth emphasizing that the SN
limit is subject to large uncertainties, mainly due to the lack of a
detailed understanding of the SN1987A. Actually it has been recently
pointed out that it might not apply at all~\cite{Bar:2019ifz}. HB
stars bounds as well as
CAST+SUMICO~\cite{Moriyama:1998kd,Inoue:2008zp,Arik:2008mq,Anastassopoulos:2017ftl}
limits can be substantially relaxed if the photon-ALP coupling and ALP
mass depend on enviromental conditions such as the matter density and
temperature, see e.g.~\cite{Jaeckel:2006xm} and references therein.

The triangular white region in Fig.~\ref{fig:gagg-sensitivities}, the
so-called \emph{cosmological triangle}, is subject as well to
cosmological constraints of which BBN and $N_\text{eff}$  are particularly
relevant~\cite{Cadamuro:2011fd,Millea:2015qra,Depta:2020wmr}. Under
standard cosmological arguments ($\Lambda$CDM) with a high
inflationary scale that region is therefore ruled out. Departures from
standard arguments, however, open this spot in parameter
space.  Possible scenarios
include low-reheating temperature ($T_R$) models, extra contributions
to the radiation energy density and non-vanishing neutrino chemical
potentials~\cite{Millea:2015qra,Depta:2020wmr}. Let us take the case
of low-reheating models. If $T_R\gg T_\text{fo}$, where $T_\text{fo}$
is the Primakoff freeze-out temperature, ALPs are abundant in the heat
bath. Photons produced by the decay of this thermal ALP distribution
will then dilute the baryon density and disrupt $N_\text{eff}$, the
former affecting BBN. If instead $T_R\ll T_\text{fo}$---and ALPs are
not produced by e.g. inflaton decays---ALPs will be much less abundant
and so the amount of photons injected in the heat
bath~\cite{Depta:2020wmr}, thus having a small effect on BBN and
$N_\text{eff}$.

From Fig.~\ref{fig:gagg-sensitivities} one can see that $90\%$CL
sensitivities will cover regions where CAST and SUMICO as well as the
SN 1987A and HB stars cooling arguments have placed limits. As we have
mentioned, these bounds can be evaded if environmental
effects---relevant in stellar interiors---are at work. Reactor
experiment measurements therefore will be able to test whether such
effects are present.  Sensitivities cover as well the cosmological
triangle, thus implying that this type of measurements will test the
cosmological hypotheses behind the constraints in that region. These
searches therefore will be complementary to those that will be carried
out at BELLE-II
($50\,\text{ab}^{-1}$)~\cite{Dolan:2017osp}. Undergoing and
near-future experiments will partially cover that region and
next-generation experiments will complement those measurements,
allowing to cover the full region . 

\begin{figure}
  \centering
  \includegraphics[scale=0.65]{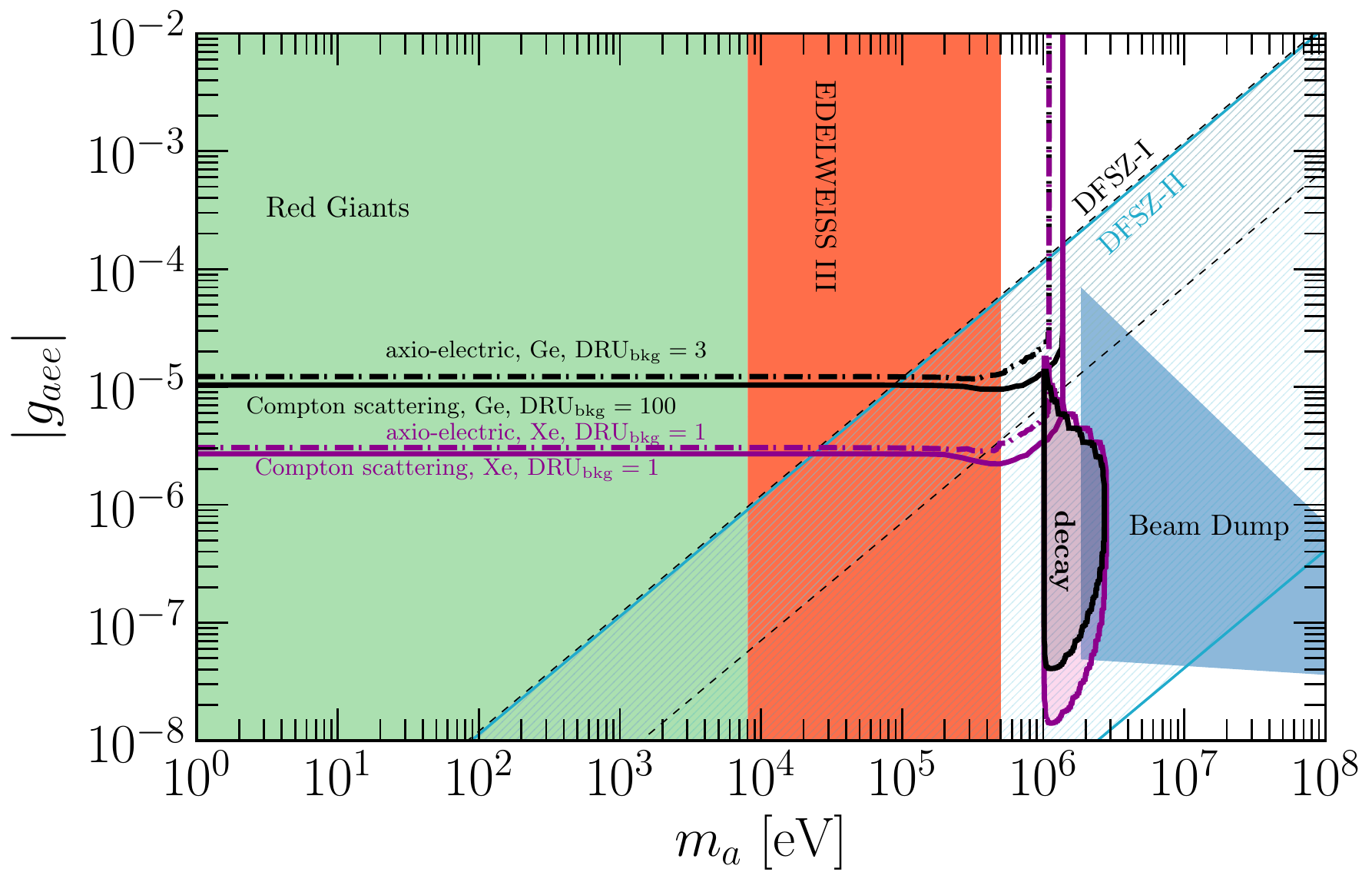}
  \caption{Electron-ALP coupling $90\%$CL sensitivities along with
    relevant laboratory and astrophysical limits that apply in the
    region of interest. Black contours correspond to $90\%$CL
    sensitivities achievable with ongoing or near-future experiments
    (see Tab.~\ref{tab:processes-and-couplings}). Purple contours
    instead indicate $90\%$CL sensitivites that could be achieved in a
    next-generation xenon detector. Solid (dashed-dotted) contours
    refer to sensitivities under the assumption of detection via inverse Compton-like scattering (axio-electric absorption) 
    and different background rates.
    See the text for more details. For completeness we show as well
    the region of DFSZ-I and DFSZ-II QCD axion models.}
  \label{fig:gaee-sensitivities}
\end{figure}
In models with dominant $g_{aee}$ coupling there are as well limits in
the region that reactor experiments can conver,
$g_{aee}\gtrsim 10^{-8}$ and in the same ALP mass range as that of the
$g_{a\gamma\gamma}$ case. In the ``low'' mass window Red Giants
cooling arguments as well as ALP searches at the EDELWEISS III
experiment place the most relevant bounds~\cite{Raffelt:2006cw,Armengaud:2018cuy}. At ``high'' ALP masses beam
dump experiments are important~\cite{Essig:2010gu}. These
limits along with the sensitivities that could be achieved in current,
near-future or next generation experiments are displayed in
Fig.~\ref{fig:gaee-sensitivities}. We show as well the region covered
by the DFSZ-I and DFSZ-II QCD axion models.

From Fig.~\ref{fig:gaee-sensitivities} one can see that the region
$m_a\subset [5,20]\times 10^5\,$eV is not subject to
constraints. Inverse Compton-like scattering and axio-electric
absorption will allow testing couplings down to $\sim 10^{-5}$
in that region, with the precise value
determined by how well backgrounds are kept under control. Smaller
values might be achievable in next generation experiments. Decay to
electron-positron pairs extends the region that can be covered in that mass
window, extending sensitivities down to couplings of order
$g_{aee} \sim 10^{-8}$. Inverse Compton-like scattering and axio-electric
absorption cover as well regions subject to limits from Red Giants
cooling arguments and EDELWEISS III. Since the latter searches for
solar ALPs, sensitivities in the region $m_a\lesssim 5\times 10^5\,$eV
will test as well possible scenarios (hypotheses) where environmental
effects could allow circumventing those limits.

\begin{figure}
  \centering
  \includegraphics[scale=0.40]{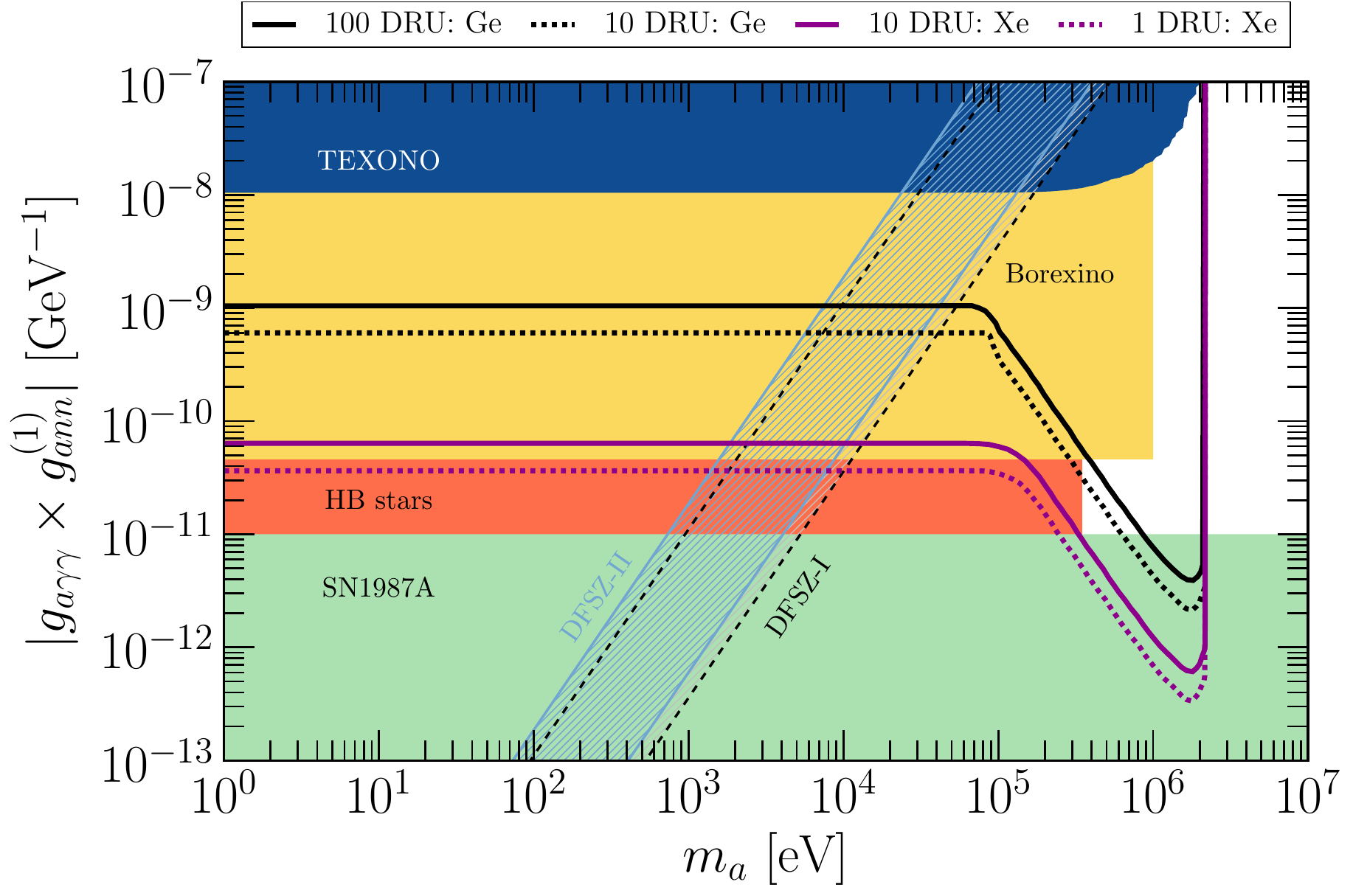}
  \includegraphics[scale=0.40]{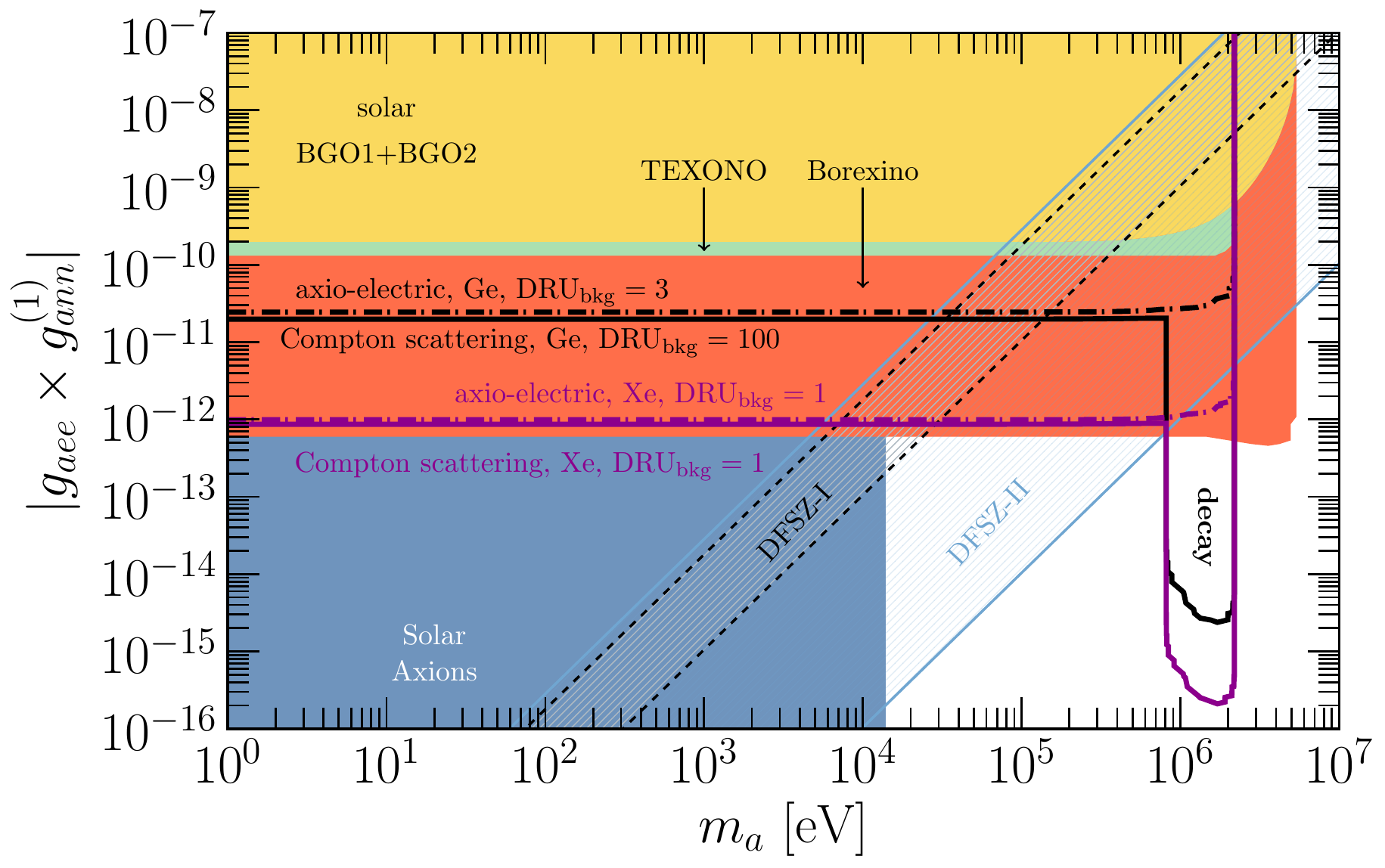}
  \caption{\textbf{Left graph}: Photon-ALP times nucleon-ALP couplings
    $90\%$CL sensitivities along with relevant laboratory and
    astrophysical limits in the region of interest. Black contours
    correspond to $90\%$CL sensitivities achievable with ongoing or
    near-future experiments (see Tab.~\ref{tab:reac_exps_det}). Purple
    contours instead indicate $90\%$CL sensitivites that could be
    achieved in a next-generation xenon detector. Solid
    (dotted) contours refer to sensitivities under the
    assumption different background rates, (see the legend). \textbf{Right graph}: Same as in
    left graph but for electron-ALP times nucleon-ALP couplings
    sensitivities. Here solid (dot-dashed) curves correspond to detection via inverse Compton-like scattering and decays into electron-positron pairs (axio-electric absorption), with different background rates. The solar axion region includes limits from
    EDELWEISS III, MAJORANA DEMONSTRATOR, CDEX and PandaX-II~\cite{Armengaud:2018cuy,Abgrall:2016tnn,Liu:2016osd,Fu:2017lfc}.  For completeness we include the region of
    DFSZ-I and DFSZ-II QCD axion models.}
  \label{fig:gaNN-sensitivities}
\end{figure}

Finally our results for ALP models with dominant (tree level)
$g_{ann}^{(1)}$ and $g_{a\gamma\gamma}$ couplings ($g_{ann}^{(1)}$ and
$g_{aee}$ couplings) are shown in the left graph (right graph) in
Fig.~\ref{fig:gaNN-sensitivities} \footnote{Due to the survival and
  decay probabilities the ALP yield has a non-trivial dependence on
  the ALP couplings. This is accounted for in our statistical analysis
  by varying the two couplings and the ALP mass independently. So
  although the plotting is done by using the product of couplings as
  the relevant variable, the results include the non-trivial coupling
  dependence.}. Relevant constraints follow from
TEXONO~\cite{Chang:2006ug},
Borexino~\cite{Bellini:2008zza,Bellini:2012kz} and
BGO~\cite{Derbin:2013zba,Derbin:2014xzr}, that cover ALP mass values
up to $10^6\,$eV. Constraints from solar ALP searches using $^{57}$Fe
first excited state and detection through axio-electric absorption
apply in the case of the $g_{aee}\times g_{ann}^{(1)}$
combination. Those limits follow from a variety of experiments, namely
EDELWEISS III~\cite{Armengaud:2018cuy}, PANDAX-II~\cite{Fu:2017lfc},
CDEX~\cite{Liu:2016osd} and the MAJORANA
DEMONSTRATOR~\cite{Abgrall:2016tnn}.  Due to kinematic constraints
they cover the ALP mass range $[1,14.4\times 10^{3}]\,\text{eV}$. In
the case of the $g_{a\gamma\gamma}\times g_{ann}^{(1)}$ combination we
include HB stars cooling and SN1987A energy loss limits, which should
be interpreted as subject to the constraints the individual couplings
are subject to,
$g_{a\gamma\gamma}\lesssim 2\times 10^{-6}\,\text{GeV}^{-1}$ (see
Fig.~\ref{fig:gagg-sensitivities}) and
$g_{ann}^{(1)}\lesssim 5\times 10^{-6}$~\cite{Lee:2018lcj}. As we have
done in the previous cases, we include as well the region covered by
the DFSZ-I and DFSZ-II QCD axion models.

\begin{figure}[ht]
\centering
\includegraphics[width = \textwidth]{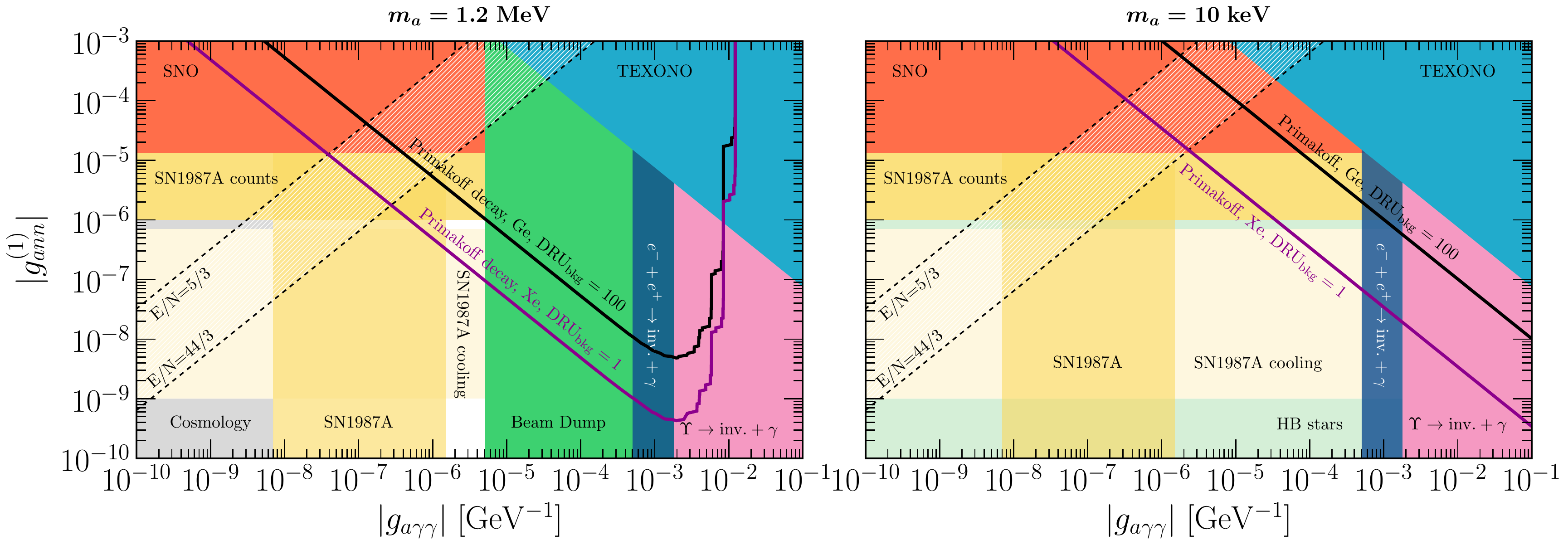}
\includegraphics[width = \textwidth]{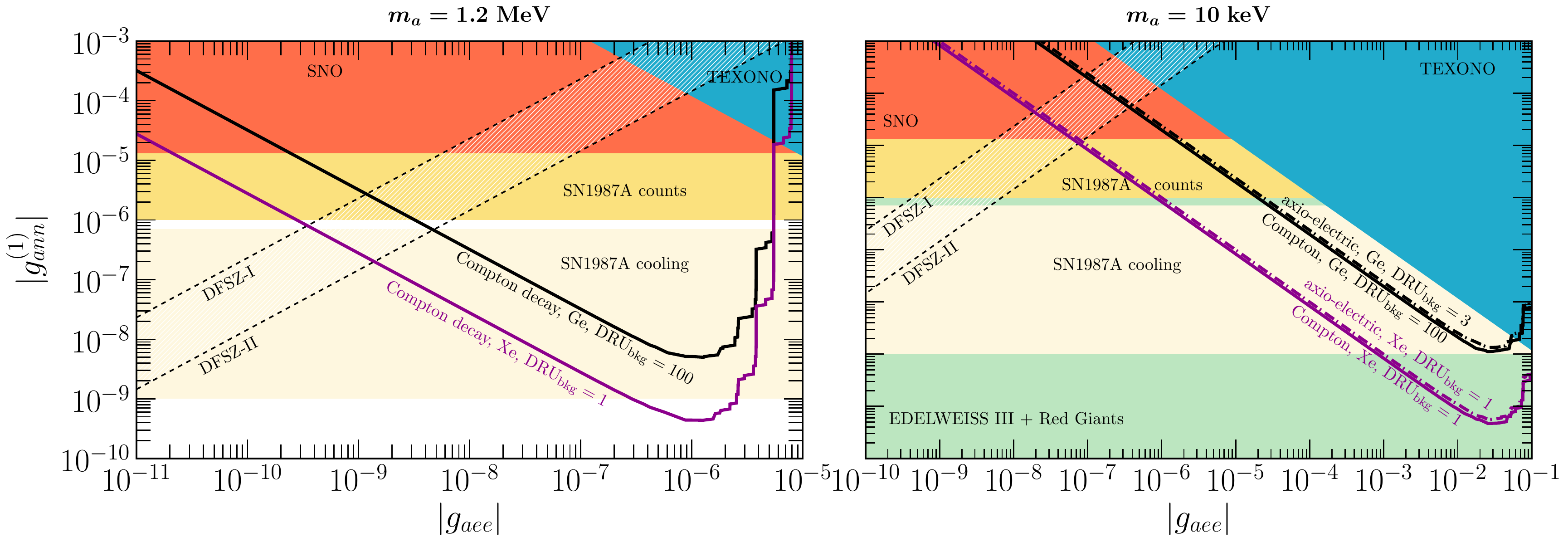}
  \caption{90\%CL sensitivities in ALP production through magnetic
    de-excitation and detection through $g_{a\gamma\gamma}$ and
    $g_{aee}$ related processes. Results are displayed in the
    couplings plane for a fixed ALP mass. See text for more details. Bounds on the isovector axion-nucleon coupling 
    $g_{ann}^{(1)}$  are adapted from~\cite{Bhusal:2020bvx}. For completeness we include the region of
    DFSZ-I and DFSZ-II QCD axion models.}
  \label{fig:couplings-planes}
\end{figure}

In the case of detection through inverse Primakoff and ALP decay to
photons, $90\%$CL sensitivity contours in the low mass range include
regions covered by TEXONO and Borexino searches. For ALP masses above
$\sim 4\times 10^5\,$eV and up to kinematic threshold
$\sim 2\times 10^6\,$eV, ALP decay will widen searches in parameter
space.  Furthermore $90\%$CL sensitivities go inside  some of the HB
stars and SN1987A regions. These measurements therefore can
partially---and potentially---test the hypotheses behind these
constraints. Detection through $g_{aee}-$related processes also cover
Borexino and TEXONO regions, but more importantly will access an ALP
mass stripe within $[8,20]\times 10^5\,$eV and down to couplings of
order $10^{-15}$. We have also shown that next-generation experiments
have the capability to substantially improve those sensitivities.

There is an infinite set of individual couplings $g_{aXX}$
($X=\gamma,e$) and $g_{ann}^{(1)}$ that satisfy the condition
$g_{aXX}\times g_{ann}^{(1)}=\text{const}$, which is not explicit in
our results in Fig.~\ref{fig:gaNN-sensitivities}. The interplay of the
individual couplings on sensitivities can be seen in
Fig.~\ref{fig:couplings-planes}, which shows the $90\%$CL
sensitivities calculated in the two-dimensional coupling plane for
fixed ALP mass along with the limits the individual couplings are
subject to.  For illustrative purposes, we have fixed $m_a = 1.2$ MeV
(10 keV) for the left (right) graphs in
Fig.~\ref{fig:couplings-planes}, depicting sensitivities for ALPs
detected via decay (scattering) processes, respectively. Indeed,
decays are more relevant at higher masses, while scattering processes
lead to almost constant number of events for masses
$m_a \lesssim 10^4~(10^5)$ eV for Primakoff (Compton-like,
axio-electric) detection mechanisms. From these plots one can see that
most of the parameter space inside these sensitivities is already
probed by a variety of different experiments, from laboratory to
astrophysics and cosmology~\cite{Bhusal:2020bvx,Calore:2020tjw}. In the case of decays however, reactor
searches could probe some small regions of parameter space currently
uncovered, at $g_{ann}^{(1)} \lesssim 10^{-9}$ and
$g_{aee} \sim 10^{-6}$, or in the ``cosmological triangle'' region at
$g_{ann}^{(1)} \sim 10^{-6}$ and
$g_{a\gamma\gamma} \sim 5\times10^{-6}$. In the rest of parameter
space, reactor searches would still give valuable information, by
providing complementary laboratory probes especially in the regions
currently tested only by astrophysical observations.

% ---------------------
% Section: Conclusions
% ---------------------
\section{Conclusions}
\label{sec:conclusions}
Nuclear reactor power plants produce a large amount of photons and so
they are suitable sources for ALPs production. In this paper we have
studied such possibility by considering production through three
possible mechanisms: Compton-like and Primakoff scattering as well as
nuclear de-excitation in the fuel material.  Motivated by the current
CE$\nu$NS experimental program, which aims at observing CE$\nu$NS
induced by reactor neutrino fluxes using various low-threshold
technologies, we have analyzed the prospects for detection of ALPs in
such experiments.  We have assumed a $10\,$kg germanium detector,
which we regard as representative of current
(CONUS~\cite{Hakenmuller:2019ecb}) or near-future CE$\nu$NS
experiments (including MINER~\cite{Agnolet:2016zir},
Ricochet~\cite{Billard:2016giu}, $\nu$GeN~\cite{Belov:2015ufh} and
TEXONO~\cite{Wong:2016lmb}).  We have also explored  future
capabilities of next generation experiments, assuming a ton-scale
xenon detector.

Concerning ALPs detection, we have considered processes controlled by
the same couplings that determine their production. Namely, we have
focused on inverse-Primakoff scattering and ALP decays into photons
for production through Primakoff scattering (processes controlled by
$g_{a\gamma\gamma}$), while we have investigated inverse Compton-like
scattering, axio-electric absorption and decay into electron pairs for
ALPs produced through Compton-like scattering (processes controlled by
$g_{aee}$). For nuclear de-excitation we have instead considered
detection through processes controlled by $g_{a\gamma\gamma}$ and
$g_{aee}$ couplings.

We have calculated $90\%$CL sensitivities to ALPs in the mass range
($1-10^7\,$eV), and we have confronted them to existing
constraints. Our main results are displayed in
Figs. \ref{fig:gagg-sensitivities}-\ref{fig:couplings-planes}, and
they can be summarized as follows. For models with dominant
$g_{a\gamma\gamma}$ coupling, these experiments will probe a region in
the ALP mass range around 1 MeV, presently constrained only by
cosmological observations.  Moreover, since reactor searches will
cover regions currently constrained by CAST+SUMICO, SN 1987A and HB stars
cooling arguments, they can also potentially test ALPs
environment-dependent properties in stellar media. For ALP models with
dominant ALP-electron coupling, reactor searches will probe ALP masses
in the range $[5,20]\times 10^5\,$eV and ALP-electron couplings
extending down to $g_{aee} \sim 10^{-8}$.  They will also cover
regions constrained by Red Giants cooling arguments as well as by
EDELWEISS III solar axions searches, thus testing also in this case
the potential relevance of ALP environmental effects.

Searches for ALPs produced in nuclear de-excitation require the
presence of two different ALP-SM couplings. If detection proceeds
through ALP-photon coupling induced-processes, they can improve
current limits set by Borexino solar axion searches and by the TEXONO
reactor experiment in the ALP mass range $\sim [0.2,2]\,$MeV. 
Moreover these searches could probe regions constrained by SN1987A
data and so the different hypotheses this limit comes along
with. Detection through ALP-electron coupling induced-processes will
cover regions already explored by other experiments such as BGO,
Borexino and TEXONO.  Still, a small region in parameter space
(currently unexplored) with ALP masses $\sim$ MeV and couplings
$g_{ann}^{(1)} \sim 10^{-9}$ could be tested.

To summarize, we have considered the possibility that ongoing and
near-future reactor-based CE$\nu$NS experiments can be used for ALP
searches. Of course these searches could be run in any other
reactor-based experiment.  However since CE$\nu$NS experiments will
run anyway, they will probably be the most suited environment for
these type of searches. We have shown that sensitivities can cover
interesting regions in the ALP parameter space not yet explored by
laboratory experiments, thus motivating the inclusion of ALP searches
in the physics program of such reactor neutrino experiments.
\section*{Acknowledgment}
We thank Bhaskar Dutta and Adrian Thompson for very useful discussions
and providing us details of their calculation in
Ref.~\cite{Dent:2019ueq}. We thank Henry Wong for useful comments and
Enrico Nardi for comments on the manuscript.  We are also very grateful to
Alessandro Mirizzi for pointing out useful updates of the astrophysical bounds 
from Globular Cluster stars and SN 1987A. DAS is supported by the
grant ``Unraveling new physics in the high-intensity and high-energy
frontiers'', Fondecyt No. 1171136.  VDR acknowledges financial support
by the SEJI/2020/016 grant (project ``Les Fosques''), funded by
Generalitat Valenciana and partial support by the Spanish grants
FPA2017-90566-REDC (Red Consolider MultiDark), FPA2017-85216-P and
PROMETEO/2018/165 (Generalitat Valenciana). LJF is supported by a
posdoctoral CONACYT grant, CONACYT CB2017-2018/A1-S-13051 (M\'exico)
and DGAPA-PAPIIT IN107118/IN107621. The work of DKP is co-financed by Greece
and the European Union (European Social Fund- ESF) through the
Operational Programme ``Human Resources Development, Education and
Lifelong Learning" in the context of the project ``Reinforcement of
Postdoctoral Researchers - 2nd Cycle" (MIS-5033021), implemented by
the State Scholarships Foundation (IKY).
\bibliography{references}

\end{document}